\newif\ifSC

\SCtrue
\ifSC
\documentclass[onecolumn,draftclsnofoot,12pt]{IEEEtran}
\else
\documentclass[final]{IEEEtran}
\fi

\usepackage{cite}
\usepackage{amsthm,amssymb,graphicx,multirow,amsmath,color,amsfonts}
\usepackage{algorithmic}
\usepackage{color}
\usepackage{subfigure}
\usepackage[update,prepend]{epstopdf}
\usepackage{bbm}
\usepackage{setspace}
\usepackage[latin1]{inputenc}

\usepackage{etoolbox}

\newtoggle{SC}
\togglefalse{SC}
\ifSC
\toggletrue{SC}
\fi

\usepackage{bbm}
\usepackage{amssymb}
\usepackage{amsmath,amsthm}
\usepackage{color}
\usepackage{amsfonts}
\usepackage{graphicx}
\usepackage{verbatim}
\usepackage{epstopdf}
\usepackage{cite}
\usepackage{tabulary}
\usepackage{mathtools}
\usepackage{enumitem}

\newcommand{\set}[1]{\mathsf{#1}}
\newcommand{\dist}[1]{\|#1\|}

\newcommand{\palmexpect}[1]{\mathbb{E}^0{\left[#1\right]}}

\newcommand{\dd}{\mathrm{d}}

\newcommand\expect[1]{\mathbb{E}\left[#1\right]}
\newcommand\prob[1]{\mathbb{P}\left[#1\right]}

\newcommand\indside[1]{\mathbbm{1}\left({#1}\right)}

\newcommand{\Ball}{\mathcal{B}}
\newcommand{\expects}[2]{\mathbb{E}_{#1}\left[#2\right] }

\newcommand{\ie}{{\em i.e. } }

\newcommand{\bigconditioned}{\left.\vphantom{\frac34}\right|}
\newcommand{\y}{{\mathbf{y}}}
\newcommand{\x}{{\mathbf{x}}}
\newcommand{\expU}[1]{e^{#1}}
\newcommand{\X}{\mathbf{x}}
\newcommand{\Y}{\mathbf{y}}
\newcommand{\expS}[1]{\exp{\left(#1\right)}}

\def\home{\hbox{\kern3pt \vbox to13pt{}%
   \pdfliteral{q 0 0 m 0 5 l 5 10 l 10 5 l 10 0 l 7 0 l 7 5 l 3 5 l 3 0 l f
               1 j 1 J -2 5 m 5 12 l 12 5 l S Q }%
   \kern 13pt}}

\newtheorem{theorem}{Theorem}

\newtheorem{corollary}{Corollary}

\newcommand{\axis}[1]{$\mathsf{#1}$}
\def\BibTeX{{\rm B\kern-.05em{\sc i\kern-.025em b}\kern-.08em
		T\kern-.1667em\lower.7ex\hbox{E}\kern-.125emX}}
\newcounter{relctr} 
\everydisplay\expandafter{\the\everydisplay\setcounter{relctr}{0}} 

\newcommand\numeq[1]%
{\stackrel{\scriptscriptstyle(\mkern-1.5mu#1\mkern-1.5mu)}{=}}
\AtBeginDocument{}

\renewcommand{\x}{\mathbf{x}}
\renewcommand{\y}{\mathbf{y}}

\newcommand{\s}{\mathbf{s}}

\newcommand{\1}{\mathbbm{1}}
\newcommand{\pt}{\mathrm{p}}
\newcommand{\drm}{\mathrm{d}}

\newcommand{\ob}{\mathrm{o}}

\newcommand{\Nt}{\mathrm{N}}
\newcommand{\Ct}{\mathrm{C}}
\newcommand{\dv}{\mathrm{d}}
\newcommand{\bt}{\mathbf{b}}
\newcommand{\B}{\mathcal{B}}

\newcommand{\kb}{\lambda_{\mathrm{d}}}
\newcommand{\st}[2]{\alpha(#1)}

\newcommand{\KS}{\mathrm{KS}}
\newcommand{\A}[3]{\mathcal{A}(#1,#2,#3)}
\renewcommand{\gg}[2]{\frac{\alpha(r)}{2 \lambda_\drm}}
\newcommand{\C}{\mathcal{C}}
\newcommand{\E}{\mathbb{E}}

\newcommand{\D}{D~}

\graphicspath{{figs/}}

\newcommand{\Phid}[1]{\Phi_{(#1),\mathrm{d}}}
\newcommand{\UBt}[1]{\overline{#1}}
\newcommand{\LBt}[1]{\underline{#1}}
\newcommand{\dime}{n}
\newcommand{\matern}{Mat\'ern~}
\newcommand{\Phip}{\Phi_\mathrm{p}}
\newcommand{\Phic}{\Phi}
\newcommand{\Yd}[2]{\Y_{(#1),#2}}
\newcommand{\Z}{\mathbf{z}}

\newcommand{\ggsq}[2]{\frac{\alpha^2{r}}{4 \lambda_\drm^2}}

\iftoggle{SC}
{}
{}

\newcommand{\commst}[1]{{\color{red}#1}}
\newcommand{\commadd}[1]{{\color{green}#1}}

\newcommand{\textadds}[1]{{\color{black}#1}}

\renewcommand{\commst}[1]{}
\renewcommand{\commadd}[1]{}
\allowdisplaybreaks 
\begin{document}

\title{\huge On the Contact and Nearest-Neighbor Distance Distributions for the $n$-Dimensional \matern Cluster Process 
}

%
%
%

\author{Kaushlendra Pandey, Harpreet S. Dhillon, Abhishek K. Gupta\,\,\vspace{-2em}
	\thanks{K. Pandey and A. K. Gupta are with IIT Kanpur, India, 208016. Email:\{kpandey,gkrabhi\}@iitk.ac.in. H. S. Dhillon is with Wireless@VT, Virginia Tech, Blacksburg, VA, USA. Email: hdhillon@vt.edu.}
} \vspace{-1em}

\maketitle

\thispagestyle{empty}
\pagestyle{empty}
\begin{abstract}
	
This letter  provides exact characterization of the contact and nearest-neighbor distance distributions for the $n$-dimensional ($n$-D) \matern cluster process (MCP). We also provide novel upper and lower bounds to these distributions  
in order to gain useful insights about their behavior. The two and  three dimensional versions of these results are directly applicable to the performance analyses of wireless networks modeled as MCP. 

\end{abstract}

\IEEEpeerreviewmaketitle

\section{Introduction}

Poisson cluster process (PCP) has recently been used as a tractable model for capturing the formation of user hot-spots in the stochastic geometry-based analyses of wireless networks\cite{afshang2016modeling,saha2017enriched,saha20183gpp,afshang2018poisson}. As is usually the case in stochastic geometry, the contact and nearest-neighbor distance distributions play a crucial role in such analyses. Recall that the contact distance (CD) is the distance of the closest point of the point process (PP) from an arbitrary point (useful in characterizing the power of the serving link in cellular networks) and the nearest neighbor distance (NND) is the distance of the nearest neighbor from a typical point of the PP (which reflects the network connectivity). While distributions of both these distances are identical for a Poisson point process (PPP), it is not so in a PCP. The focus of this letter  is on characterizing these distributions for an MCP, a special case of PCP, which has  recently received attention because of its relevance in modeling user hot-spots (very similar models have also been used in 3GPP simulations \cite{saha20183gpp}). 

The CD and NND distributions of an MCP were first characterized for the 2-D case in \cite{afshang2017nearest} (and for Thomas cluster process, another special case of PCP, in \cite{afshang2016nearest}). However, the expressions involve multiple integrals and are thus unwieldy. A simpler expression for the CD for a 2-D MCP was derived in \cite{azimi2018stochastic}. In general, there are two main approaches to these derivations. The first one uses the probability generating functional (PGFL) of a PCP \cite{moyal1962general} and is also the approach taken in \cite{afshang2017nearest}. \textadds{ For a detailed discussion on MCP, readers are advised to refer \cite{afshang2017nearest} and references therein.} The second approach is to condition on the parent PPP of the PCP and then use the PGFL
of a PPP. While this general approach has been used recently for the coverage analysis of cellular networks in \cite{saha2019unified}, \cite{miyoshi2018downlink}, we will demonstrate in this letter  that a similar approach can also be leveraged to derive remarkably simple expressions for the CD and NND distributions of an  $n$-D MCP. We also develop novel closed-form upper and lower bounds on these distributions in order to provide further insights. While the construction of these bounds is seemingly straightforward (involves developing simple bounds on the intersection of $n$-D balls), the resulting bounds are remarkably tight, which is also verified using numerical comparisons.

\textbf{Notation:}  $\Ball(\y,r)$ denotes an $\dime$-\D ball of radius $r$ centered at location $\y$.  $\Phi(\set{A})$ denotes the number of points of $\Phi$  in set $\set{A}$. Let $\A{r}{r_\drm}{x}$ be the volume of intersection between two $\dime$-\D balls with radii $r_\drm$ and $r$ that are $x$ distance apart. Let 
$v_\dime$ denote the volume of unit $\dime$-\D ball. Let $\x+\set{A}$ denote  set obtained by shifting each point in $\set{A}$ by vector $\x$. 
Let $\beta(r)=\min\{r,r_\drm\}$ and $C(n,i)=
\binom{n}{i}$.

\section{\matern Cluster Process}
We first define the cluster process (CP). Let $\Phip=\{\X_i,i\in\mathbb{N}\}$ be a PP where $\X_i$ denotes the location of the $i$-th point. For each point $\X_i$, associate a PP 
$\Phid{i}=
\{
\Yd{i}{j}
\}
$. The point $\X_i$ is termed the {\em parent point} and $\Phid{i}$ is its {\em daughter point process}. The first PP, $\Phip$, consisting of all parent points is called the {\em parent process} \cite{chiu2013stochastic}. The union $\Phi$ of all daughter PPs centered at their parent points $\X_i$ is termed as CP \ie
\begin{align*}
\Phic&=\cup_{\X_i\in\Phip}\X_i+\Phid{i}\\
&=\left\{\mathbf{z}_{ij}:\mathbf{z}_{ij}=\X_i+\Yd{i}{j},\X_i\in\Phip,\Yd{i}{j}\in\Phid{i}\ \forall \ \ i,j\right\}.
\end{align*}

In this letter, we consider a stationary MCP which is a special case of CP that satisfies the following properties: 
\begin{enumerate}
	\item The parent PP is a PPP with density $\lambda_\pt$.
	\item Each daughter PP is a  PPP with density $\lambda_\drm$ in the ball $\Ball(\ob,r_\drm)$ with center at the origin and radius $r_\drm$.
	\item Daughter PPs are placed at the parent points independently of  each other and of the  parent process.
\end{enumerate} 
 Note that the average number of points in each daughter PP $\Phid{i}$ is $m=\lambda_\drm\pi r_\drm^2$.
In this letter, we are interested
in the distributions of the two random variables: CD $R_{\Ct}$, and NND $R_{\Nt}$, associated with the
MCP.
\section{Contact Distance Distribution}
Recall that the CD is the distance of the closest point of $\Phi$ from an arbitrary point, which can be placed at the origin because of the stationarity of $\Phi$. We start the derivation by noting that the event that $R_{\Ct}$ is greater than $r$ is equivalent to the event that there is no point with distance  less than $r$ from the origin. In other words,
\begin{align}
F_{R_\Ct}(r)&=1-\prob{R_\Ct>r}=1-\prob{\min_{\Z\in\Phi}\dist{\Z} >r }\nonumber \\
&=1-\prob{\dist{\Z} >r\ \forall \ \Z\in\Phi },
\end{align}
which is equal to the void probability of set $\Ball(\ob,r)$. The CD distribution for the MCP  is given in the following theorem (See Appendix \ref{app:thm:1} for the proof.).

\begin{theorem}
	\label{thm:1}
	The CDF of CD 
	of an $\dime$-\D MCP is
	 \iftoggle{SC}{}{$F_{R_\Ct}(r)=$}
	\begin{align}
	\iftoggle{SC}{F_{R_\Ct}(r)=}{}
	&1-\expS{-v_\dime\dime \lambda_{\pt}
	\int_{0}^{r+r_\drm}\left(1-e^{-\kb \A{r}{r_\drm}{x}}\right)x^{\dime-1}  \dv x}	\nonumber\\
		=&1-\exp\left(-v_\dime \lambda_{\pt} 
	\left((r+r_\drm)^{\dime}
-|r-r_\drm|^\dime \expS{-\lambda_\drm v_\dime \beta^\dime(r)}\right.\right. \nonumber \\
&\left.\left. -n\int_{|r-r_\drm|}^{r+r_\drm}(\expS{-\kb \A{r}{r_\drm}{x}}
x^{\dime-1}  \dv x)
\right)
\right)\label{eq:FinalMCDN}.
\end{align}
\end{theorem}
\noindent\textbf{Special Cases}:
\begin{enumerate}
\item For $1$-D MCP ($n=1$), $\A{r}{r_\drm}{x}=\min\{r+r_\drm-x,2\beta(r)\}$. Hence, CD distribution is given as
\begin{align}
&F_{R_\Ct}(r)=1-\exp\left(-2{\lambda_\pt}\left( 
(r+r_\drm)\vphantom{e^{-jjjj_k}}-|r-r_\drm|\expU{-\lambda_\drm2\beta(r)}
+(\expU{-\lambda_\drm2\beta(r)}-1)/{\lambda_\drm}
\right)\right).\label{eq:1Dexact}
\end{align}
\item For the  2-\D ($\dime=2$) case, $\A{r}{r_\drm}{x}=$
\begin{align}	
\hspace{-.3in} \begin{cases}
r_{\drm}^2\cos^{-1}\left(\frac{x^2+r_\drm^2- r^2}{2xr_{\drm}}\right)
+r^2\cos^{-1}\left(\frac{x^2+ r^2-r_\drm^2}{2xr}\right)& \iftoggle{SC}{}{\text{If\,} r+r_\drm \ge }\\ 
-\frac12
\sqrt{((r_\drm+r)^2-x^2)(x^2-(r_\drm-r)^2)},
&
\hspace{-.1in}\iftoggle{SC}{\text{if } r+r_\drm \ge}{ }\, x \ge |r-r_\drm|\\
\pi\min(r,r_\drm)^2=\pi\beta^2(r)&\text{\,\,\,\,\,\,otherwise} 
\end{cases}\nonumber
\end{align}
The closed form expression may not be possible for this case. However, the derived CDF for CD  \eqref{eq:FinalMCDN} is significantly simpler than the one in \cite[Eq.~4]{afshang2017nearest}.
\end{enumerate}
\subsection{Bounds on $F_{R_\Ct}(r)$}
Since, it may not be possible to derive closed form expression
for some cases, we next provide two sets of closed form upper and
lower bounds for the same in the next two theorems.
	
\begin{theorem} \label{thm:UBLB1ContactDistance}
The CDF of contact distance of $n$-\D MCP is  upper and lower  bounded respectively as 
\begin{align}
\iftoggle{SC}{&\UBt{F_{R_{\Ct}}}(r)}{&\UBt{F_{R_{\Ct}}}(r)}=1-
\exp\left[
-v_\dime  \lambda_{\pt}
\left(
\vphantom{\eta^{\frac1\alpha}}
{(r+r_\drm)^n}
-{|r-r_\drm|}^{n}\expU{-\lambda_\drm v_\dime\beta^{\dime}(r)}-n!\sum_{k=0}^{n-1}
\frac{(-1)^{n-k-1}}{(k!)}
{\left(\lambda_\drm 2^{n-1}\beta^{n-1}(r)\right)}^{k-n}
\right.\right.\nonumber \\
 &\left.\left.\left({\left(r+r_\drm\right)}^k -
{\left|r-r_\drm\right|}^k\expU{-2^n\lambda_\drm \beta^{n}(r)}
\right)
\right)
\right].\label{eq:UB1}\\
\iftoggle{SC}{&\LBt{F_{R_{\Ct}}}(r)=}{&\LBt{F_{R_{\Ct}}}(r)=}1-
	\exp\left[
	-v_\dime  \lambda_{\pt}
	\left(
		\vphantom{\eta^{\frac1\alpha}}
	{(r+r_\drm)^n}
	-{|r-r_\drm|}^{n}\expU
	{
	-\lambda_\drm v_\dime\beta^{\dime}(r)
	}
	-
	{(r+r_\drm)}^{n}\sum_{i=0}^{n-1}
	C({n-1},{i}){(-1)}^i2^{i+1}\times
	\right.\right.\nonumber\\
&\left.\left.
	{(\lambda_\drm v_\dime(r+r_\drm)^n)}^{-(i+1)/{n}}\gamma\left(
	({i+1})/{n},\lambda_\drm v_\dime \beta^\dime (r)
	\right)
	\right)
	\right],\label{eq:LB11}
\end{align}
where $\gamma(s,x)=\int_{0}^{x}t^{s-1}e^{-t}\drm t$, is incomplete gamma function. 
\begin{IEEEproof}
	See Appendix \ref{app:thm:2}.
	\end{IEEEproof}
\end{theorem}

\begin{theorem} \label{thm:UBLB2ContactDistance}
The upper and  lower bound on the contact distance distribution $F_{R_\Ct}(r)$ are given as:
	\begin{align*}
	\UBt{\UBt{F_{R_\Ct}}}(r)=&
	1-
		\exp\left(
			-v_\dime\lambda_\pt 
		\left( (r+r_\drm)^\dime
		\left(
		1-e^{-v_\dime\lambda_\drm\beta^\dime(r)}
		\right)
		\right)\right),
	\\
	\underline{\underline{F_{R_\Ct}}}(r)=&1-\exp\left(-v_n \lambda_\pt |r-r_\drm|^n
	\left(1-e^{-\lambda_\drm v_n\beta^n(r)}\right)
	 \right).
	\end{align*} 
	Proof: For the upper bound and lower bounds, we replace $\A{r}{r_\drm}{x}$ respectively by its upper bound $v_\dime \lambda_\drm \beta^\dime(r)
	\indside{r+r_\drm\le x}$ and lower bound 0 in \eqref{eq:FinalMCDN}. 
\end{theorem}
\subsection{Asymptotic behavior of $F_{R_\Ct}(r)$ with $r_\drm$}
\textbf{Case-I: $r_\drm\rightarrow 0$}:  

As $r_\drm\rightarrow 0$, both the upper and lower bounds given in Theorem \ref{thm:UBLB1ContactDistance} converge to the function $g_0(r)=1-\exp\left(-v_n\lambda_\pt r^n(1-e^{-m})\right)$. \textadds{Using the squeeze theorem \cite[Th. 3.3.6]{sohrab2003basic}, we can show that $F_{R_\Ct}(r)$  also converges to $g_0(r)$.}

Note that $g_0(r)$ is the distribution of a PPP with intensity $\lambda_{\pt}(1-e^{-m})$. This convergence can be understood in the following way. As $r_\drm\rightarrow0$,  all daughter points of a parent point become co-located at the location of that parent point. The number of points co-located at any parent point is distributed as \textsf{\small Poisson}($m$).
This means that some parent points may not have any daughter point resulting in the absence of point at these sites. Hence, the density of sites that have at-least one point would be  $\lambda_0=\lambda(1-e^{-m})$. 
Note that the resultant PP  is not a PPP, but a multi-set with site locations distributed as PPP($\lambda_0$) and each site $\mathbf{s}$ having $m_\mathbf{s}$ points co-located at it. 
\textbf{Case-II: $r_\drm\rightarrow\infty$}: 
Here, $\beta(r)=r$.
As $r_\drm\rightarrow \infty$, both bounds given in Theorem \ref{thm:UBLB2ContactDistance} converge to the function $g_\infty(r)=1-\exp(-mv_n\lambda_{\pt}r^n)$. Hence,  $F_{R_\Ct}(r)$  also converges to $g_\infty(r)$. 

\section{{Nearest neighbor  distance distribution}} 
Since the MCP process is stationary, the typical point $\Z'$ of the point process can be taken at the origin without loss of generality. 
Now the event that  $R_{\Nt}$ is greater than $r$ is equivalent to the event that there is no other point in $\Phic$ with distance less than $r$ from the typical point $\Z'$. In other words,
\begin{align*}
F_{R_{\Nt}}(r)&=1-\prob{\Phic|\Ball(\ob,r)|=1\bigconditioned\Z'=\ob \in \Phic}\\
&=1-\mathbb{P}_{\ob}^{!}\left(\Phic|\Ball(\ob,r)|)=0\right).
\end{align*} 
Here, the $\mathbb{P}_{\ob}^{!}$ is the reduced palm distribution. Solving further, as discussed in Appendix \ref{app:thm:4}, we get the following Theorem.  
\begin{theorem}
	\label{thm:ExactNearestDistance}
The CDF of the nearest neighbor distance $F_{R_{\Nt}}(r)$, for the  $\dime$-\D MCP is (See Appendix \ref{app:thm:4} for proof)
\begin{align}\label{eq:NND}
=1-(1-F_{R_\Ct}(r)){\dime}{r_{\drm}^{-\dime}} \int_{0}^{r_{\drm}}e^{
	-\kb
	\A{r}{r_\drm}{x}}
x^{\dime-1}\dd x.
\end{align}
\end{theorem} 
 
\begin{corollary}
	If $r>2r_{\drm}$,  $\B(\x,r)$ will  cover $\B(\ob,r_{\drm}) \ \ \forall\x$. Hence, $\A{r}{r_\drm}{x}=v_\dime r_\drm^{\dime}$. So, for $r>2r_{\drm}$, \eqref{eq:NND} simplifies to 
	\begin{align} 
	\label{closedformMCDN}
	F_{R_{\Nt}}(r)&
	=1-\left(1-F_{R_\Ct}(r)\right)e^{-m}.
	\end{align}
\end{corollary}

\begin{theorem}
	\label{thm:UBLB1NearestDistance}
For $r\leq 2 r_\drm$, the upper and lower bounds on the CDF of NND for the $n$-D MCP is (See Appendix \ref{ndnearestneighbor} for proof):
	\begin{align*}
\overline{F_{R_{\Nt}}}(r)=&1-{(1-\overline{F_{R_\Ct}}(r))}{r_\drm^{-n}}\left[\vphantom{\frac1\Delta_n}e^{-\lambda_{\drm}v_n \beta^n(r)}{|r-r_\drm|^n}\right.\\&\left.+\sum_{k=0}^{n-1}(-1)^{n-1-k}{(n)!}/{k!}(\lambda_{\drm}2^{n-1}\beta^{n-1}(r))^{k-n}\right.\nonumber\\
&\left.\left[e^{-\lambda_{\drm}2^{n-1}\beta^{n-1}(r) r}r_\drm^{k}-e^{-\lambda_{\drm}2^{n}\beta^{n}(r)}|r-r_\drm|^{k}\right]\right],\\
\underline{{F_{R_\Nt}}}(r)=&1-{(1-\underline{F_{R_\Ct}}(r))}{r_\drm^{-n}}
\left[
\vphantom{\frac1\Delta_n}
e^{-\lambda_{\drm}v_n \beta^n(r)}{|r-r_\drm|^n}\right.\\
&\left.+ \left[\vphantom{\frac1\Delta_n}\sum_{i=0}^{n-1}C(n-1,i)(-1)^{i}(r+r_\drm)^{n-1-i}((\lambda_{\drm}v_n)^{-1} 2^{n})^{\frac{i+1}{n}}\right.\right.\nonumber\\
&\left.\left.\left[\gamma({(i+1)}/{n}, \lambda_{\drm} v_n \beta^n(r))-\gamma({(i+1)}/{n},\lambda_{\drm}v_n 2^{-n} r^n)\right]\vphantom{\frac1\Delta_n}\right]\right].
\end{align*}
\text{For $r>2r_\drm$ the upper and lower bound on NND is:}
\begin{align*}
&\overline{F_{R_{\Nt}}}(r)=1-(1-\overline{F_{R_\Ct}}(r))e^{-m},\\	
&\underline{F_{R_{\Nt}}}(r)=1-(1-\underline{F_{R_\Ct}}(r))e^{-m}.
\end{align*}
\end{theorem}
\textadds{Another upper bound over $F_{R_\Nt}(r)$ can be obtained as
	\begin{align*}
	\overline{\overline{F_{R_{\Nt}}}}(r)&=1-{(1-{\overline{F_{R_\Ct}}}(r))}\left(1-e^{-\lambda_{\drm}v_n \beta^n(r)}\right).
	\end{align*}}
 \begin{figure}
	\centering
	\includegraphics[width=.6\columnwidth]{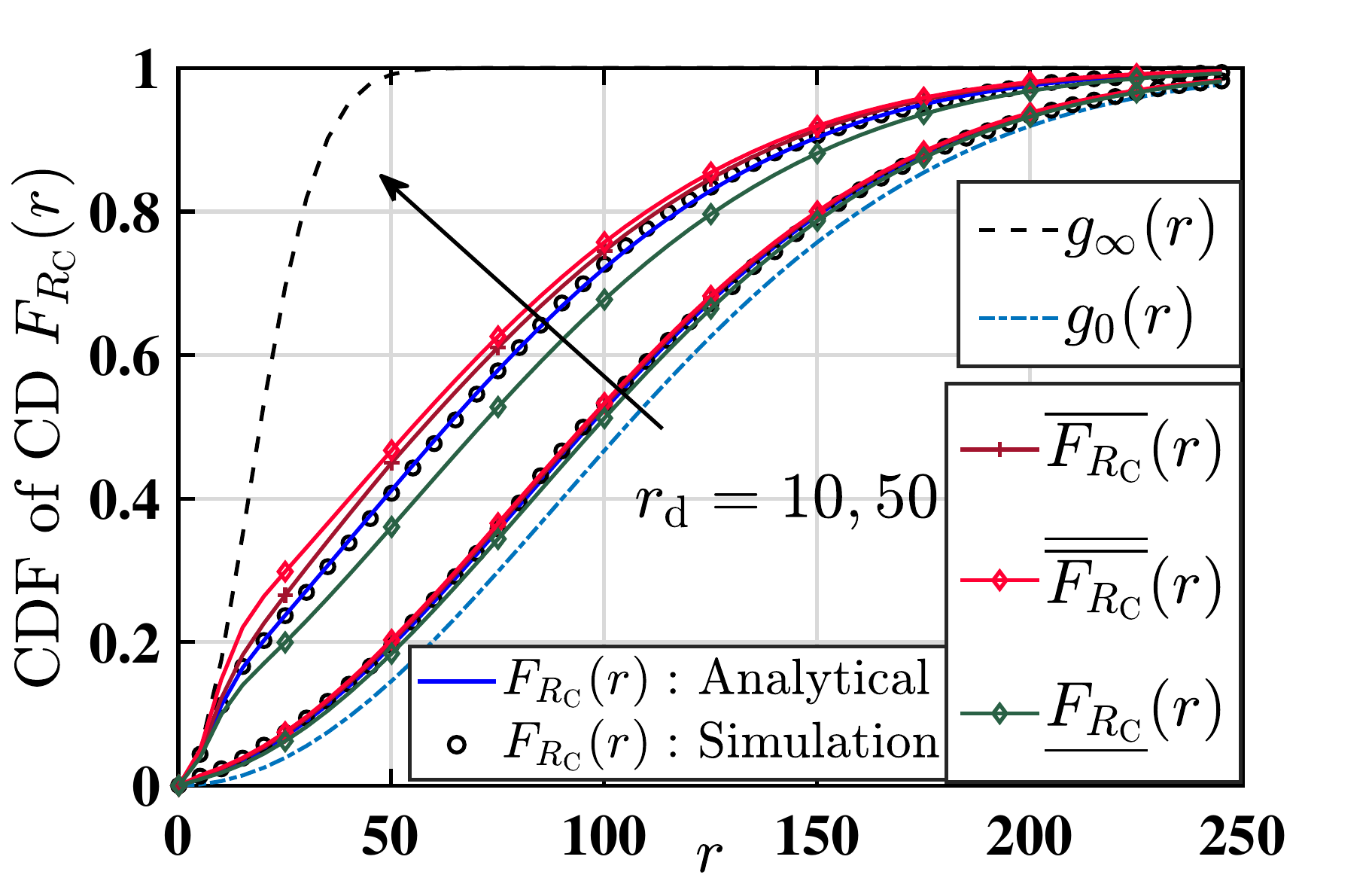}\\
	\includegraphics[width=.6\columnwidth]{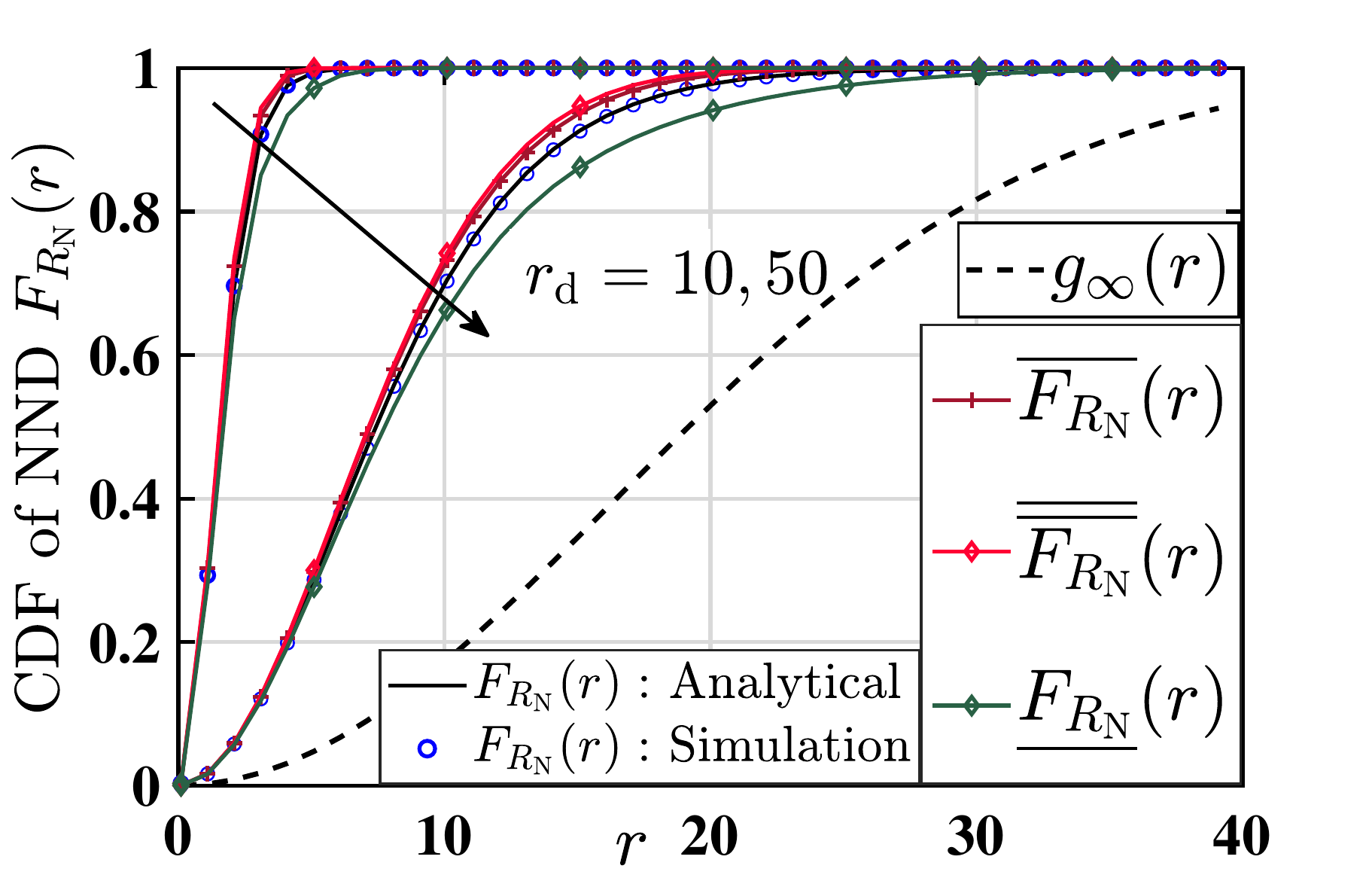}
	\caption{CDF  and bounds of CD and NND ($m=30, \lambda_{\pt}=20\times10^{-6}$).}
	\label{CCDF4}
\end{figure}
Fig. \ref{CCDF4} shows the CDF of $R_{\Ct}$ and $R_\Nt$ for 2D MCP and their corresponding bounds. We notice that the CD increases with $r_\drm$. 
\section{Conclusions}
In this letter, we have characterized the exact CDFs of the CD and NND for the $n$-D MCP. Our expressions are much simpler and compact when compared to their 2-D counterparts available in the literature. By constructing simple bounds on the intersection of two $n$-D balls, we also derived remarkably tight closed-form upper and lower bonds on these CDFs. The 2-D and 3-D
versions of our results have plenty of applications in the performance analyses of a variety of  clustered wireless networks. \textadds{ The general $n$-D results obtained in this letter can be specialized to emerging 3D deployment scenarios, such as dense urban deployments \cite{7421294}, and deployments in multi-floor malls/stadiums.
} 
\renewcommand{\bt}{\Ball}

\appendices

\section{}
\label{app:thm:1} 
The void probability  of MCP  is 
 \begin{align*}
\mathbb{P}(\Phi|\bt(\ob,r)|=0)&=\mathbb{E}\left[\prod_{\X_i \in \Phip}\prod_{\Yd{i}{j}\in \Phid{i}}\indside{\left(\X_i+\Yd{i}{j}\right)\notin \bt(\ob,r)}\right]\\
&=\mathbb{E}_{\Phip}\left[\prod_{\X_i\in\Phip}\mathbb{E}_{\Yd{i}{j}|\X_i}\left[\prod_{
	{\Yd{i}{j}\in {\bt(\ob,r_{\drm})}}
}
 \hspace{-.15in}
 \indside{(\X_i+\Yd{i}{j})\notin \bt(\ob,r)}
\right]\right] \\
&=\mathbb{E}_{\Phip}\left[\prod_{\X_i\in\Phip}\mathbb{E}_{\Yd{i}{j}|\X_i}\left[\prod_{\Yd{i}{j}\in \bt(\ob,r_{\drm})}
\indside{\Yd{i}{j}\notin \bt(-\X_i,r)}
\right]\right] \\
&\stackrel{(a)}=\mathbb{E}_{\Phip}
\left[
	\prod_{\X_i\in\Phip}
	\expS{-\lambda_\drm  |\bt(\ob,r_\drm)\cap\bt(-\X_i,r)|}
\right] \\
&=\exp\left(-\lambda_{\pt}\int_{\mathbb{R}^n}\left(1-\exp\left(-\kb\int_{\s\in\B^{\x}}(\1\{(\x+\s)\in \bt(0,r)\})\dv\s\right)\right)\dv\x\right)\\
&=\exp\left(-\lambda_{\pt}\int_{\mathbb{R}^n}\left(1-\exp\left(-\kb\int_{\s \in \bt(0,r_{\drm})}(\1\{\s \in \bt(-\x,r)\})\dv\s\right)\right)\dv\x\right)\\
&\stackrel{(b)}=\exp\left(-\lambda_{\pt}\int_{\mathbb{R}^n}\left(1-\expS{
	-\kb \left|\bt(\ob,r_{\drm})\cap \bt(-\x,r)\right|
}\right)\dv\x\right).
\end{align*}
\noindent \hspace*{-.09in} Here, $(a)$ is due to the void probability of PPPs $\Phid{i}$ and $(b)$ is due to the PGFL of  $\Phip$ \cite{AndGupDhi16}. Using the expression for the volume of the  intersection of two balls, we get Theorem \ref{thm:1}.
\renewcommand{\st}{\alpha}
\begin{figure}[ht!]
	\def\svgwidth{\textwidth}
	\centering
	\includegraphics[width=.8\textwidth]{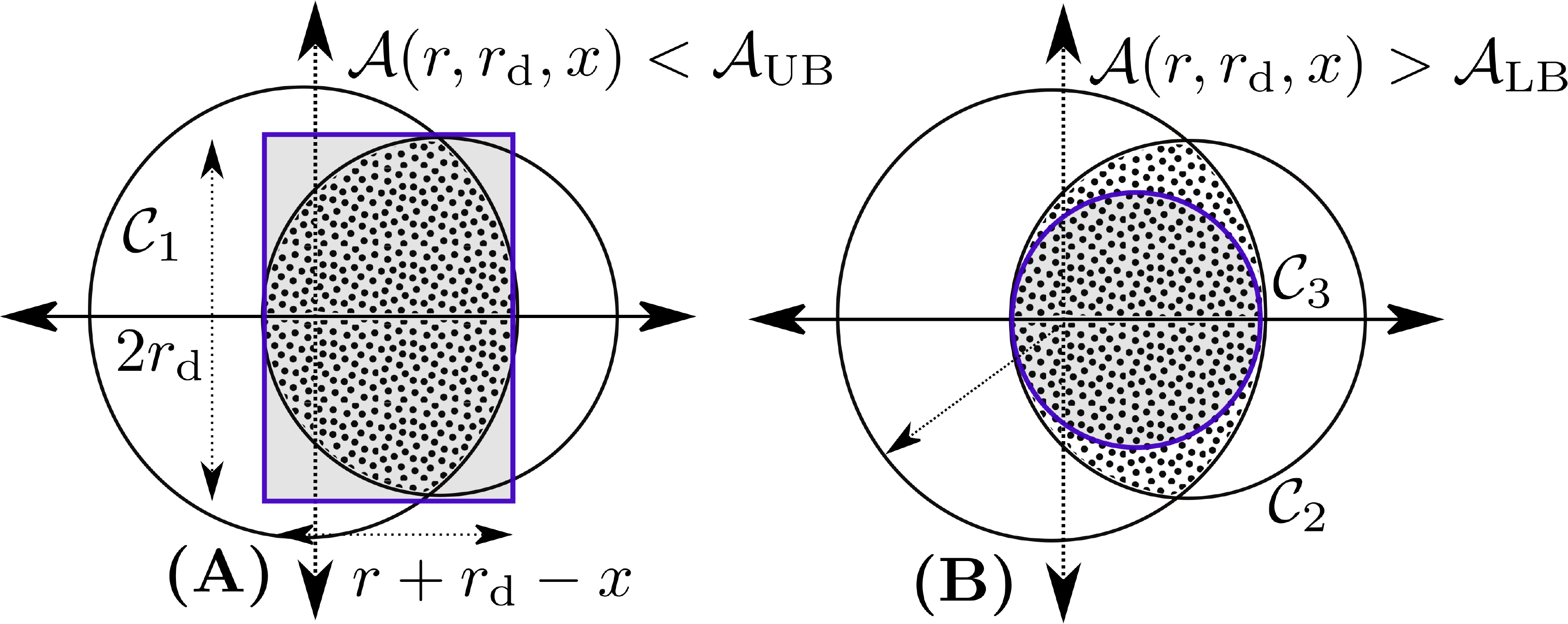}
	\caption{(A) A hyperrectangle  with \axis{x}-length $r+r_{\drm}-x$ and the other lengths $2\min(r,r_{\drm})$  will cover the entire intersection region of the two balls $\C_1$ and $\C_2$. (B) A ball of radius $\frac{r+r_{\drm}-x}{2}$ will have a volume smaller than the volume of intersection of two balls $\A{r}{r_\drm}{x}$. Although the illustration is for $2$-D, the idea works for general $n$-D.  } \label{fig:UBLB}
\end{figure}
\section{}\label{app:thm:2}
Let us focus on the last integral in \eqref{eq:FinalMCDN}, which is
	\begin{align}
	I=n\int_{|r-r_\drm|}^{r+r_\drm}
	e^{-\kb \A{r}{r_\drm}{x}}
	x^{\dime-1}  \dv x
	\label{eq:Idef1}.
	\end{align}
We will bound the area $\A{r}{r_\drm}{x}$ (dotted area in Fig. \ref{fig:UBLB}) which is the area of intersection of the balls $\C_1:\Ball(\ob,r)$ and $\C_2:\Ball(\x,r_\drm)$ with $\dist{\x}=x$ with $|r-r_\drm|<x<r+r_\drm)$. Without loss of generality, take  $\x$ on the \axis{x}-axis.
For the upper bound, we note that the intersection has \axis{x}-length $r+r_\drm-x$ and maximum length $2\min{(r,r_\drm)}$ in the rest of the dimensions. It can be  contained inside the hyperrectangular region with volume $(r+r_\drm-x){(2\min(r,r_\drm))}^{\dime-1}$ (see shaded area in Fig. \ref{fig:UBLB}(A)). Using this upper bound in \eqref{eq:Idef1}, we get

	\begin{align*}
	I\ge n&
	\int_{|r-r_\drm|}^{r+r_\drm}\expS{-\lambda_\drm    
		2^{n-1}\beta^{n-1}(r)(r+r_\drm-x)}x^{n-1}\dv x\\
	=&n!
	\sum_0^{n-1}{(-1)^{n-k}}/{k!}
	{\left(\lambda_\drm 2^{n-1}\beta^{n-1}(r)\right)}^{k-n}
	\left(
	{\left(r+r_\drm\right)}^k
	-{\left|r-r_\drm\right|}^k
	\expS{-2^n\lambda_\drm \beta^{n}(r)}
	\right).
	\end{align*}
Substituting this bound in \eqref{eq:FinalMCDN}, we get the desired upper bound \eqref{eq:UB1}.
For the lower bound, let us first assume $r>r_\drm$ without loss of generality. Construct a third ball $\C_3:\Ball(({(r+r_\drm-x)}/{2},\textbf{0}_{n-1}),\frac12(r+r_\drm-x))$. Since, $\C_3$ touches both balls $\C_1$ and $\C_2$ on a single points and these two points lie inside the intersection of $\C_1$ and $\C_2$, it must lie inside the intersection region. Hence, $\A{r}{r_\drm}{x}$ is greater than the volume of $\C_3$ which is $v_\dime { (r+r_\drm-x)^\dime}/{2^\dime}$. Using
this lower bound in \eqref{eq:Idef1}, we get
	\begin{align*}
	&I\le n{\int_{|r-r_\drm|}^{r+r_\drm}e^{-{\lambda_\drm v_n}{2^{-n}}(r+r_\drm-x)^n}x^{n-1}\dv x}
	\\
	&=\int_{0}^{\lambda_\drm v_\dime\beta^\dime(r) }
	e^{-y}
	\left(r+r_\drm-(
	\frac{2y}
	{\lambda_\drm v_\dime })^{\frac{1}{n}}
	\right)^{n-1}
	{(\lambda_\drm v_\dime)^\frac{-1}{n}}
	y^{\frac{1}{n}-1}\dv y.
	\end{align*}
\noindent where the last step is due to the substitution $\lambda_\drm v_\dime (r+r_\drm-x)^n=y$. 
Now using binomial expansion of $(a+b)^n$ and then using the definition of
$\gamma(.)$
 we get the desired bound \eqref{eq:LB11}.

\newcommand{\obb}{0}
\newcommand{\typoint}{\Z^{(\obb)}_\ob}
\renewcommand{\typoint}{\Z'}
\section{}
\label{app:thm:4} 
 
Let typical point $\Z'$ belong to the daughter point process $\Phid{\obb}$ and let $\X_{\obb}$ be the parent point of $\Phid{\obb}$. 
Since $\Z'$ is located at the origin $\ob$, $\Phid{\obb}$ is a finite PPP in $\B(\ob,r_{\drm})$. It can be shown that the parent point $\X_{\ob}$ of $\Z_i$ is uniformly distributed  in $\B(\ob,r_{\drm})$.
It follows trivially that $\typoint\in\X_{\obb}+\Phid{\obb}$. Therefore, the CCDF of NND of this typical point can be expressed as: 
\begin{align} 
&F_{R_\Nt}(r)=F_{R_\Nt(\typoint)}(r)=\mathbb{P}^{\ob}\left[R_\Nt(\typoint)\le r\right]\\
&=1-\prob{R_\Nt(\typoint)>r|\typoint=\ob\in\Phic}\\
&=1-{\E^{\ob} \left[\prod_{\X_i\in\Phip}\prod_{\Yd{i}{j} \in \Phid{i} \setminus \{\ob\}}\indside{\dist{\X_i+\Yd{i}{j}}>r} \right]}\\
&\stackrel{(a)}{=}\expects{\X_0}{\E^{\X_0} \left[\prod_{\X_i\in\Phip}
	\expect{\prod_{\Yd{i}{j} \in \Psi'}\indside{\dist{\X_i+\Yd{i}{j}}>r}}\right]}\nonumber\\
&\stackrel{(b)}{=}\expects{\X_0}{
		\palmexpect{\prod_{
				{\Y_{(0),j}\in}{  \Psi
			}
		}\indside{\dist{\X_0+\Yd{i}{j}}>r}}
	\E^{!\X_0} \left[
	\prod_{\X_i}
	g(\X_i)
	 \right]}\nonumber
\end{align}

where 
$\Psi=\Phid{0} \setminus \{\ob\}$ and

	\begin{align*}
	g(\X_i)=\expect{\prod\nolimits_{\Yd{i}{j} \in \Phid{i} }\indside{\dist{\X_i+\Yd{i}{j}}>r}}.
	\end{align*}

Here $(a)$ is due to the total probability law and $(b)$ is due to the independence of $\Phid{0}$ and the rest of the cluster process $\Phic$. Now, from the Slivnyak theorem \cite{AndGupDhi16}, we know that $\E^{!\x}[\ ]=\E[\ ]$. Hence,

	\begin{align}
F_{R_\Nt}(r)&=1-\expects{\X_0}{\palmexpect
		{\prod_{\Yd{i}{j} \in \Psi}\hspace{-0.1in}\indside{\dist{\X_0+\Yd{i}{j}}>r}}
	}
	\E\left[
	\prod_{\X_i}
	g(\X_i)
	\right].\label{eq:eq11}
	\end{align} 

The second product term is the probability that no point of $\Phip$ is closer than distance $r$ from the origin which is equal to $1-F_{R_\Ct}(r)$. The first term can be further written as

	\begin{align}
	&\expects{\X_0}{\palmexpect
		{\prod_{\Yd{i}{j} \in \Phid{0} \setminus \{\ob\}}\indside{\dist{\X_0+\Yd{i}{j}}>r}}}\nonumber\\
	&=\expects{\X_0}{\E^{(!0)}\left[
		{\prod_{\Yd{i}{j} \in \Phid{0} }\indside{{\X_0+\Yd{i}{j}}\notin\Ball(\ob,r)}}\right]}\nonumber\\
	&=\expects{\X_0}{
		\expS{
			-\int_{\Ball(0,r_\drm)}\lambda \indside{{\y}\in\Ball(-\X_0,r)}\dd\y
	}}\nonumber\\
	&=\expects{\X_0}{
		\expS{
			-\lambda |\Ball(-\X_0,r)\cap\Ball(\ob,r_\drm)|
	}}\label{eq:secondterm}.
	\end{align}

Using the distribution of $\X_0$ 
in \eqref{eq:secondterm} and substituting the final expression in \eqref{eq:eq11} along with the expression for the second term, we arrive at the desired result.  
\section{}\label{ndnearestneighbor}
	
	 The integration interval in \eqref{eq:NND} can be broken into two parts: Interval $x\in [0,|r-r_\drm|]$ where intersection volume is $v_n \lambda_\drm \beta^n(r)$ and 

interval  $x\in[|r-r_\drm|,r_\drm]$ where we will use upper and lower bounds of $\A{}{}{}$.
Consider the integral:
	\begin{align}\label{eq:appndtwoparts}
	I'=\int_{|r-r_\drm|}^{r_\drm}e^{-	\lambda_{\drm}\A{r}{r_\drm}{x}}x^{n-1}\dv x.
	\end{align}
Substituting $\A{r}{r_\drm}{x}$ with its upper bound (derived in Appendix \ref{app:thm:2}), we get 
	\begin{align*}
	&I'\geq\int_{|r-r_\drm|}^{r_\drm}\exp(-\lambda_{\drm}2^{n-1}\beta^{n-1}(r)(r+r_\drm-x))x^{n-1}\dv x\\
	&\geq\exp(-\lambda_{\drm}2^{n-1}\beta^{n-1}(r)(r+r_\drm))\int_{|r-r_\drm|}^{r_\drm}\exp(\lambda_{\drm}2^{n-1}\beta^{n-1}(r)x)x^{n-1}\dv x.\\
	&\text{Substitute}\, t=\lambda_{\drm}2^{n-1}\beta^{n-1}(r)x, \,\text{in the above integral to get}\\
	&I\geq\left(\exp(-\lambda_{\drm}2^{n-1}\beta^{n-1}(r)(r+r_\drm))(\lambda_{\drm}2^{n-1}\beta^{n-1}(r))^{-n}\int_{\lambda_{\drm}2^{n-1}\beta^{n-1}(r)|r-r_\drm|}^{\lambda_{\drm}2^{n-1}\beta^{n-1}(r)r_\drm}\exp(t)t^{n-1}\dv t\right)\\
	&\geq\exp(-\lambda_{\drm}2^{n-1}\beta^{n-1}(r)(r+r_\drm))(\lambda_{\drm}2^{n-1}\beta^{n-1}(r))^{-n}(n-1)!\left[\sum_0^{n-1}\frac{(-1)^k}{k!}t^{k}e^t\right]_{\lambda_{\drm}2^{n-1}\beta^{n-1}(r)|r-r_\drm|}^{\lambda_{\drm}2^{n-1}\beta^{n-1}(r)r_\drm}\\
	&\geq\sum_{k=0}^{n-1}(-1)^{n-1-k}{(n)!}/{k!}(\lambda_{\drm}2^{n-1}\beta^{n-1}(r))^{k-n}\left[e^{-\lambda_{\drm}2^{n-1}\beta^{n-1}(r) r}r_\drm^{k}-e^{-\lambda_{\drm}2^{n}\beta^{n}(r)}|r-r_\drm|^{k}\right]
	\end{align*} 

Substituting the above along with the upper bound for CD distribution in \eqref{eq:appndtwoparts}, we get the desired upper bound.
Similarly for  the lower bound, we substitute the lower bounds of $\A{r}{r_\drm}{x}$ (derived in Appendix \ref{app:thm:2}) in its place to get
	\begin{align*}
	I'\leq n\int_{|r-r_\drm|}^{r_\drm}e^{-\frac{ v_n}{2^n}\lambda_{\drm}{(r+r_\drm-x)}^{n}}
	x^{n-1}\dv x.
	\end{align*}
Let $\delta={(\lambda_\drm v_n)}/{2^n},a(r)=|r-r_\drm|, b(r)=r_\drm$, $\delta{(r+r_\drm-x)}^{n}=y$ and $\kappa=2^n \delta \beta^n(r)$. Then, using binomial expansion and definition of $\gamma(.)$ we get
	\begin{align*}
	f_1(x)&=\int_{\delta r^n}^{\kappa}e^{-y}(b(r)-(\frac{y}{\delta})^{\frac{1}{n}})^{n-1}\frac{1}{\delta^{\frac{1}{n}}}y^{\frac{1}{n}-1} \dv y\\
	&=\frac{1}{n\delta^{\frac{1}{n}}}\sum_{i=0}^{n-1}\binom{n-1}{i}(r_\drm)^{n-1-i}(-1)^{i}(\frac{1}{\delta})^{\frac{i}{n}}\int_{\delta r^n}^{\kappa}e^{-y}y^{\frac{i+1}{n}-1}\dv y\\
	&=\frac{1}{n\delta^{\frac{1}{n}}}\sum_{i=0}^{n-1}\binom{n-1}{i}(r_\drm)^{n-1-i}(-1)^{i}(\frac{1}{\delta})^{\frac{i}{n}}\left[\gamma(\frac{i+1}{n}-1,\kappa)-\gamma(\frac{i+1}{n}-1,\delta r^n)\right]\\
	I'&\leq \left[\sum_{i=0}^{n-1}C({n-1},{i})(-1)^{i}(r+r_\drm)^{n-1-i}((\lambda_{\drm}v_n)^{-1} 2^{n})^{\frac{i+1}{n}}\right.\\
	&\left.\left[\gamma({(i+1)}/{n}, \lambda_{\drm} v_n \beta^n(r))-\gamma({(i+1)}/{n},\lambda_{\drm}v_n 2^{-n} r^n)\right]                                                                          \right].
	\end{align*}

Substituting the above bounds on $I'$ along with the upper/lower bounds (see Appendix \ref{app:thm:1} for the proof) on $F_{R_\Ct}(r)$ in \eqref{eq:appndtwoparts}, we get the desired  bounds.
\section{On the Goodness of Derived Bounds}
We first identify key parameters that may impact the tightness of the bounds. Using these insights, we will then compute the error between the bounds and the exact value in order to quantify the tightness of the bounds.

\textbf{Identification of Key Parameters}: 
Looking at the analytical expressions for the CDF of the contact distance and the nearest neighbor distance, we can see that the expression depends on the following three key parameters:\\
\textbf{(1)} The mean number of daughter points $m$, allocated to each parent.\\
\textbf{(2)} The radius $r_\drm$ of the ball centered at each parent location.\\
\textbf{(3)} The parent point process density $\lambda_\pt$.\\
Upon close observation of \eqref{eq:FinalMCDN}, \eqref{eq:UB1} and \eqref{eq:LB11}, we can see that both the exact expression and the bounds remain invariant if $\lambda_{\pt}$ is increased by a factor of $k^n$ and $r_\drm$ and $r$ are scaled down by a factor of $k$, that is
\begin{align}
F_{R_\Ct}(r,r_\drm,\lambda_{\pt},m)&=F_{R_\Ct}(r/k,r_\drm/k,\lambda_{\pt}k^n,m)\nonumber\\
\overline{ F_{R_\Ct}}(r,r_\drm,\lambda_{\pt},m)&=\overline{F_{R_\Ct}}(r/k,r_\drm/k,\lambda_{\pt}k^n,m)\nonumber\\
\underline{F_{R_\Ct}}(r,r_\drm,\lambda_{\pt},m)&=\underline{F_{R_\Ct}}(r/k,r_\drm/k,\lambda_{\pt}k^n,m)\label{eq:scl}
\end{align}
It is well known that, such scaling laws are quite common in the stochastic geometry literature and have also found many applications in wireless networks. For instance, the impact of such scaling laws on the performance of a fairly general 2D cellular network model was thoroughly studied in \cite{afshang2018equi}. 
Not surprisingly, the same observation can be made for the NND, as well. These results indicate that it suffices to consider only two parameters to study the tightness of the bounds. We will show this behavior with the help of numerical results. First, we define three metrics that will quantitatively evaluate the error in bounds.\\
{\textbf{(A) Kolmogorov-Smirnov (K-S) distance:}}\\
The K-S distance between the bounds and the exact expression for the contact distance distribution is given as
\begin{align*}
{D}^{\KS}_{C,U}(r_\drm,\lambda_\pt,m)&=\max_{r}|F_{R_{\Ct}}(r)-\overline{F_{R_{\Ct}}}(r)|\\
{D}^{\KS}_{C,L}(r_\drm,\lambda_\pt,m)&=\max_{r}|F_{R_{\Ct}}(r)-\underline{F_{R_{\Ct}}}(r)|.
\end{align*}
Here, ${D}_{C,U}$  and ${D}_{C,L}$ denote the deviation of upper and lower bound for the contact distance distribution.  Applying the scaling law \ref{eq:scl}, we can observe that 
\begin{align*}
{D}^{\KS}_{C,U}(r_\drm/k,\lambda_\pt k^n,m)&={D}^{\KS}_{C,U}(r_\drm,\lambda_\pt,m)\\
{D}^{\KS}_{C,L}(r_\drm/k,\lambda_\pt k^n,m)&={D}^{\KS}_{C,L}(r_\drm,\lambda_\pt,m).
\end{align*}
This shows that the deviation ${D}^{\KS}_{\cdot,\cdot}$ which was observed for some value of $\lambda_\pt$ and $r_\drm$ would also be observed for different values of the two parameters (specifically, $\lambda_\pt'$ and $r_\drm/(\lambda_\pt'/\lambda_\pt)^{1/n}$). The same deviation behavior would be observed at different values of $\lambda_\pt$, as long as the value of $r_\drm$ is appropriately scaled. . 
In particular, the maximum of ${D}^{\KS}_{C,U}(r_\drm,\lambda_\pt,m)$ over $r_\drm$ remains the same regardless of the value of $\lambda_\pt$. Therefore, it suffices to study the deviation for a specific value of $\lambda_\pt$ (without loss of generality).

Similarly, for nearest neighbor distance, maximum deviation of bounds can be defined as
\begin{align*}
{D}^{\KS}_{N,U}(r_\drm,\lambda_\pt,m)&=\max_{r}|F_{R_{\Nt}}(r)-\overline{F_{R_{\Nt}}}(r)|\\
{D}^{\KS}_{N,L}(r_\drm,\lambda_\pt,m)&=\max_{r}|F_{R_{\Nt}}(r)-\underline{F_{R_{\Nt}}}(r)|.
\end{align*}
\par
\textbf{{(B) Mean deviation:}} 
The expected value of deviation for CDFs of the contact and nearest neighbor distance is defined as:
\begin{align*}
{D}^{\mathrm{Avg}}_{C,U}(r_\drm,\lambda_\pt,m)&=\mathbb{E}\left[|F_{R_{\Ct}}(r)-\overline{F_{R_{\Ct}}}(r)|\right]=\int_{0}^{\infty}|F_{R_{\Ct}}(r)-\overline{F_{R_{\Ct}}}(r)|f_{R_{\Ct}}(r)\dv r,\\
{D}^{\mathrm{Avg}}_{C,L}(r_\drm,\lambda_\pt,m)&=\mathbb{E}\left[|F_{R_{\Ct}}(r)-\underline{F_{R_{\Ct}}}(r)|\right]=\int_{0}^{\infty}|F_{R_{\Ct}}(r)-\underline{F_{R_{\Ct}}}(r)|f_{R_{\Ct}}(r)\dv r,\\
{D}^{\mathrm{Avg}}_{N,U}(r_\drm,\lambda_\pt,m)&=\mathbb{E}\left[|F_{R_{\Nt}}(r)-\overline{F_{R_{\Nt}}}(r)|\right]=\int_{0}^{\infty}|F_{R_{\Nt}}(r)-\overline{F_{R_{\Nt}}}(r)|f_{R_{\Nt}}(r)\dv r,\\
{D}^{\mathrm{Avg}}_{N,L}(r_\drm,\lambda_\pt,m)&=\mathbb{E}\left[|F_{R_{\Nt}}(r)-\underline{F_{R_{\Nt}}}(r)|\right]=\int_{0}^{\infty}|F_{R_{\Nt}}(r)-\underline{F_{R_{\Nt}}}(r)|f_{R_{\Nt}}(r)\dv r,
\end{align*}
where $f_{R_\Ct}(r)$ and $f_{R_\Nt}(r)$ are the PDFs (probability density functions) of contact and nearest neighbor distance respectively. We can show that the similar scaling laws apply here also.

\textbf{{(C) Kullback-Leibler (K-L) Distance:}}
K-L distance is used to measure the difference between the two probability distributions defined over the same sample space. The K-L distance is $0$ if the two random variable have the same distribution. For contact and nearest neighbor distance we measure the K-L distance of bounds from the original PDF for several combination of $m$ and a wide range of $r_\drm$. The expression of K-L distance of bounds of contact and nearest neighbor distance from the original PDF is:
\begin{align*}
{D}^{\mathrm{K-L}}_{C,U}(r_\drm,\lambda_\pt,m)&=\int_{0}^{\infty}f_{R_\Ct}(r)\log\left(\frac{f_{R_\Ct}(r)}{\underline{f_{R_\Ct}}(r)}\right)\dv r,\\
{D}^{\mathrm{K-L}}_{C,L}(r_\drm,\lambda_\pt,m)&=\int_{0}^{\infty}{f_{R_\Ct}}(r)\log\left(\frac{{f_{R_{\Ct}}}(r)}{\overline{f_{R_\Ct}}(r)}\right)\dv r,\\
{D}^{\mathrm{K-L}}_{N,U}(r_\drm,\lambda_\pt,m)&=\int_{0}^{\infty}f_{R_\Nt}(r)\log\left(\frac{f_{R_\Nt}(r)}{\underline{f_{R_\Nt}}(r)}\right)\dv r,\\
{D}^{\mathrm{K-L}}_{N,L}(r_\drm,\lambda_\pt,m)&=\int_{0}^{\infty}{f_{R_\Nt}}(r)\log\left(\frac{{f_{R_{\Nt}}}(r)}{\overline{f_{R_\Nt}}(r)}\right)\dv r,
\end{align*}
where $f_{R_\Ct}(r)$ and $f_{R_\Nt}(r)$ are the PDFs of contact and nearest neighbor distance respectively.\\

\begin{figure}[ht!]
	\subfigure[K-S distance of the upper bound]{\includegraphics[width=8cm]{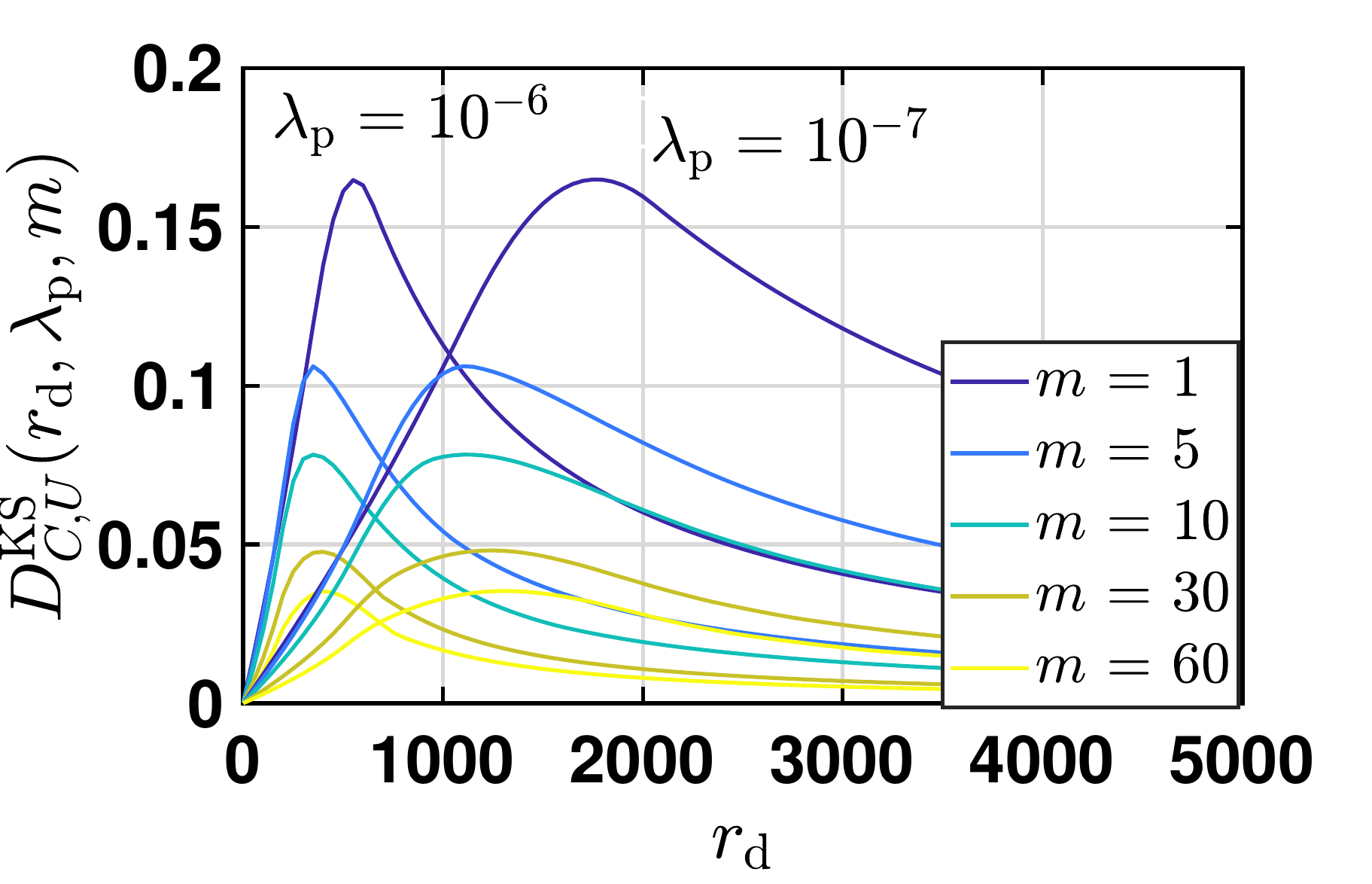}}
	\hfill
	\subfigure[KS distance of the lower bound]{\includegraphics[width=8cm]{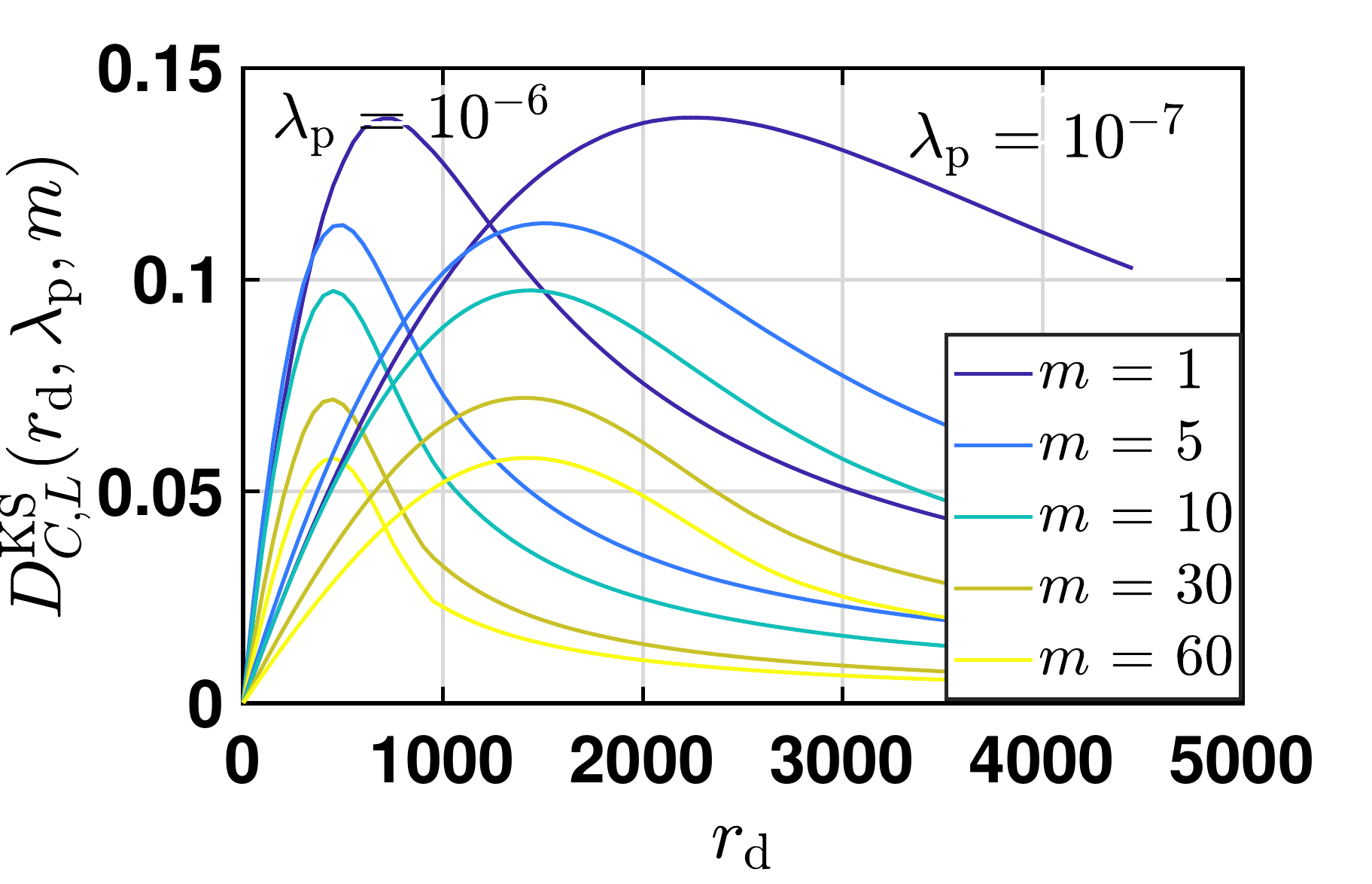}}
	\hfill
	\subfigure[Average deviation of the upper bound]{\includegraphics[width=8cm]{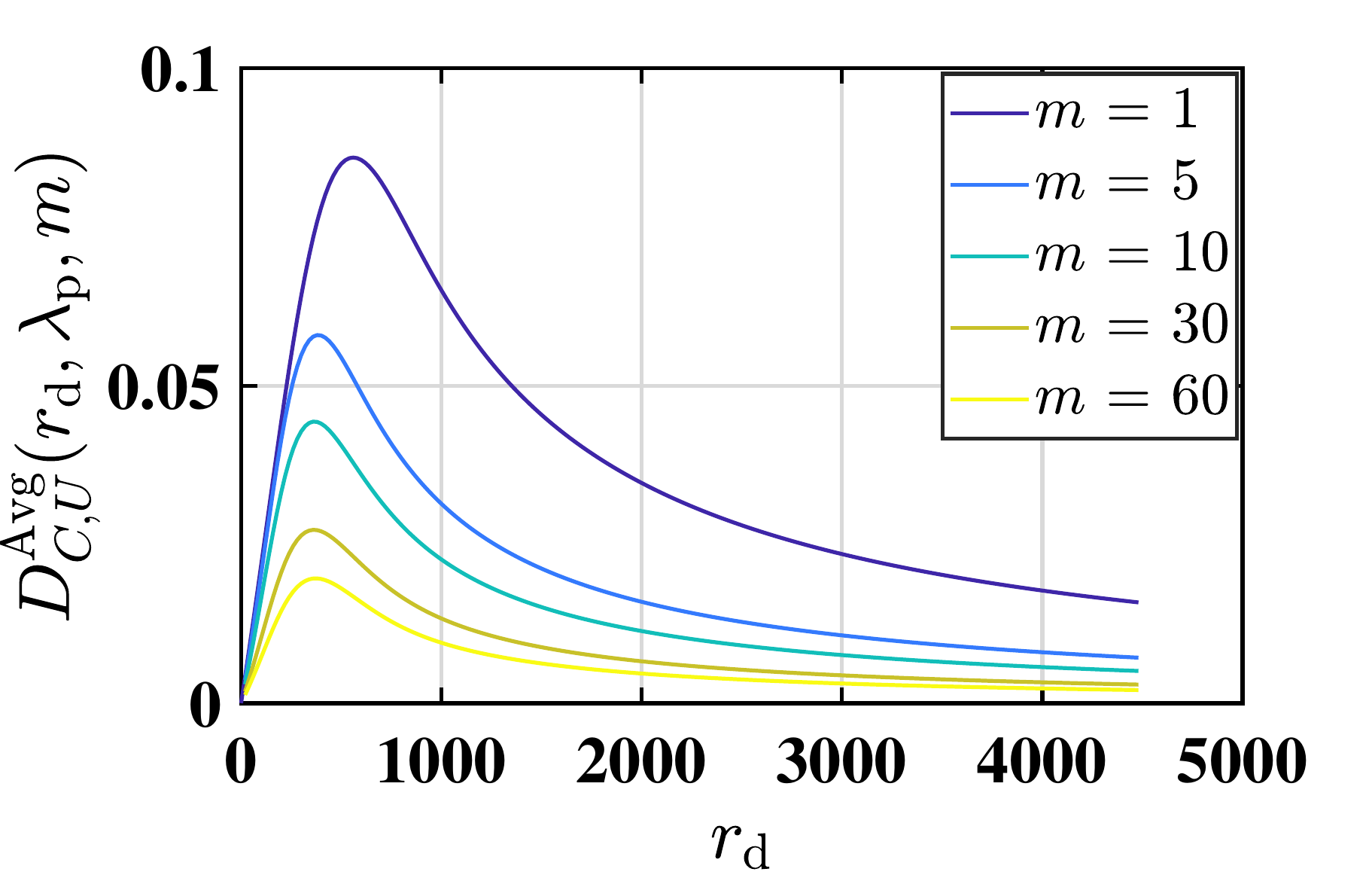}}
	\hfill
	\subfigure[Average deviation of the lower bound]{\includegraphics[width=8cm]{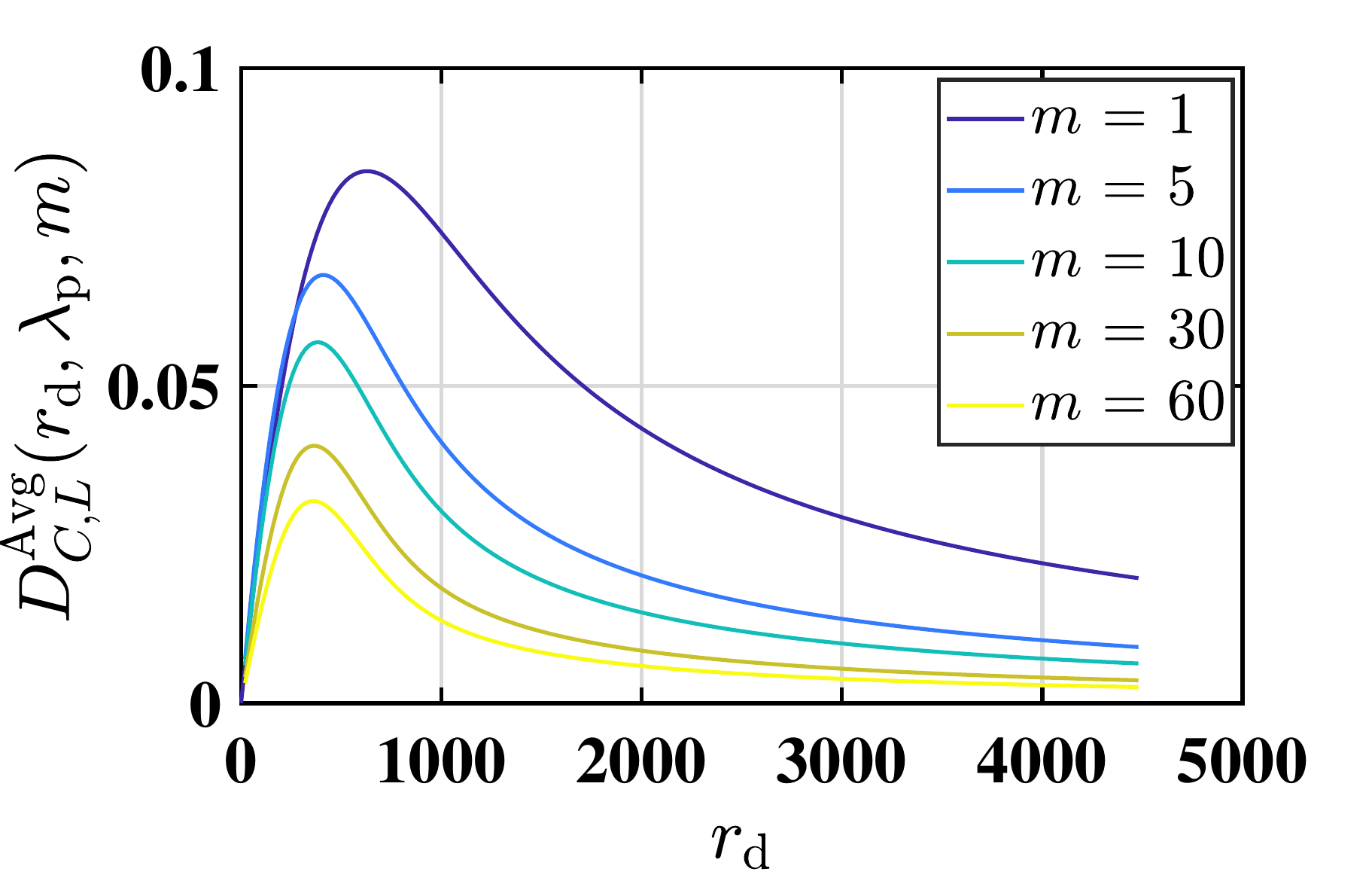}}
	\hfill
	\subfigure[K-L distance of the upper bound]{\includegraphics[width=8cm]{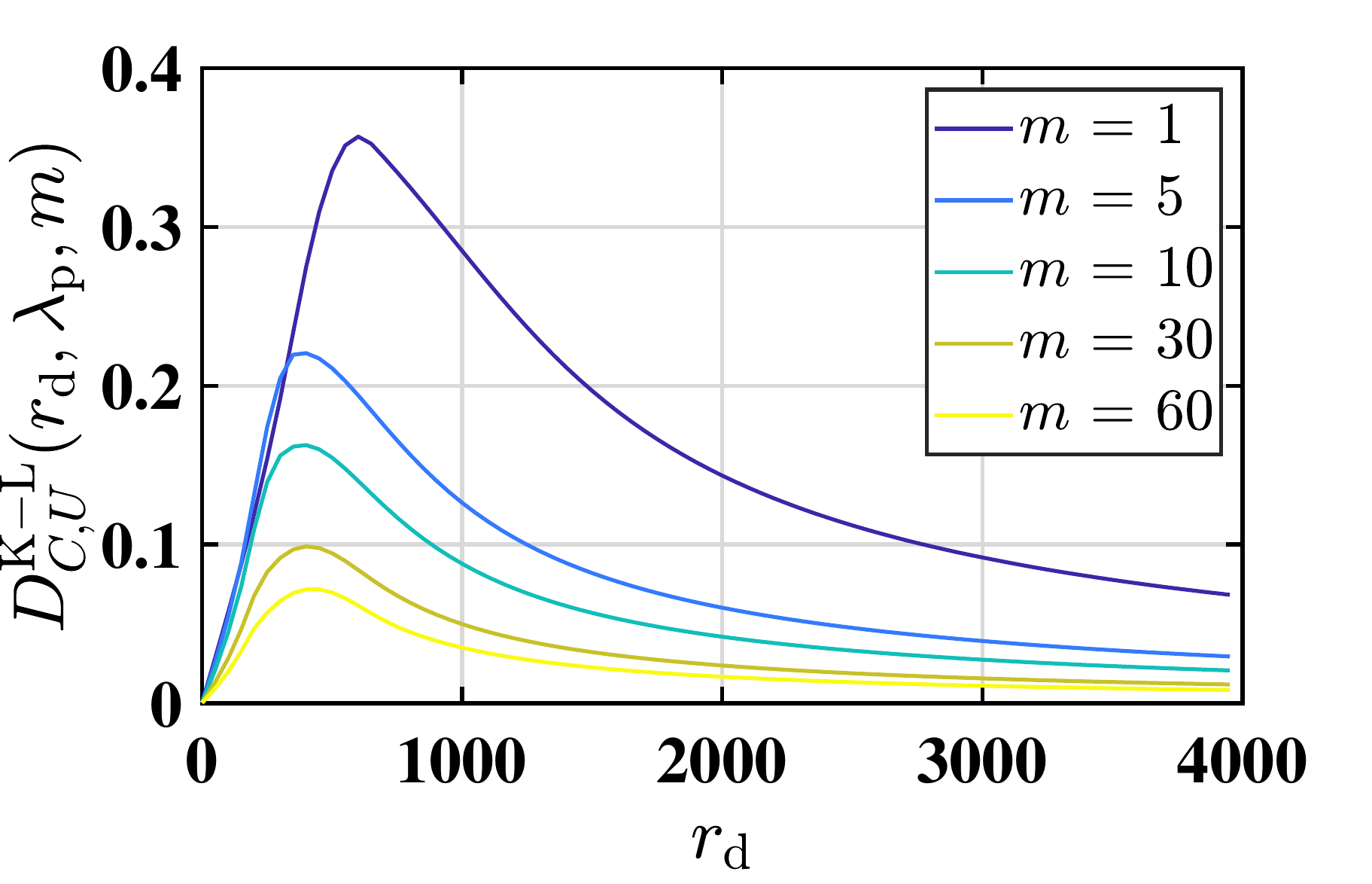}}
	\hfill
	\subfigure[K-L distance of the lower bound]{\includegraphics[width=8cm]{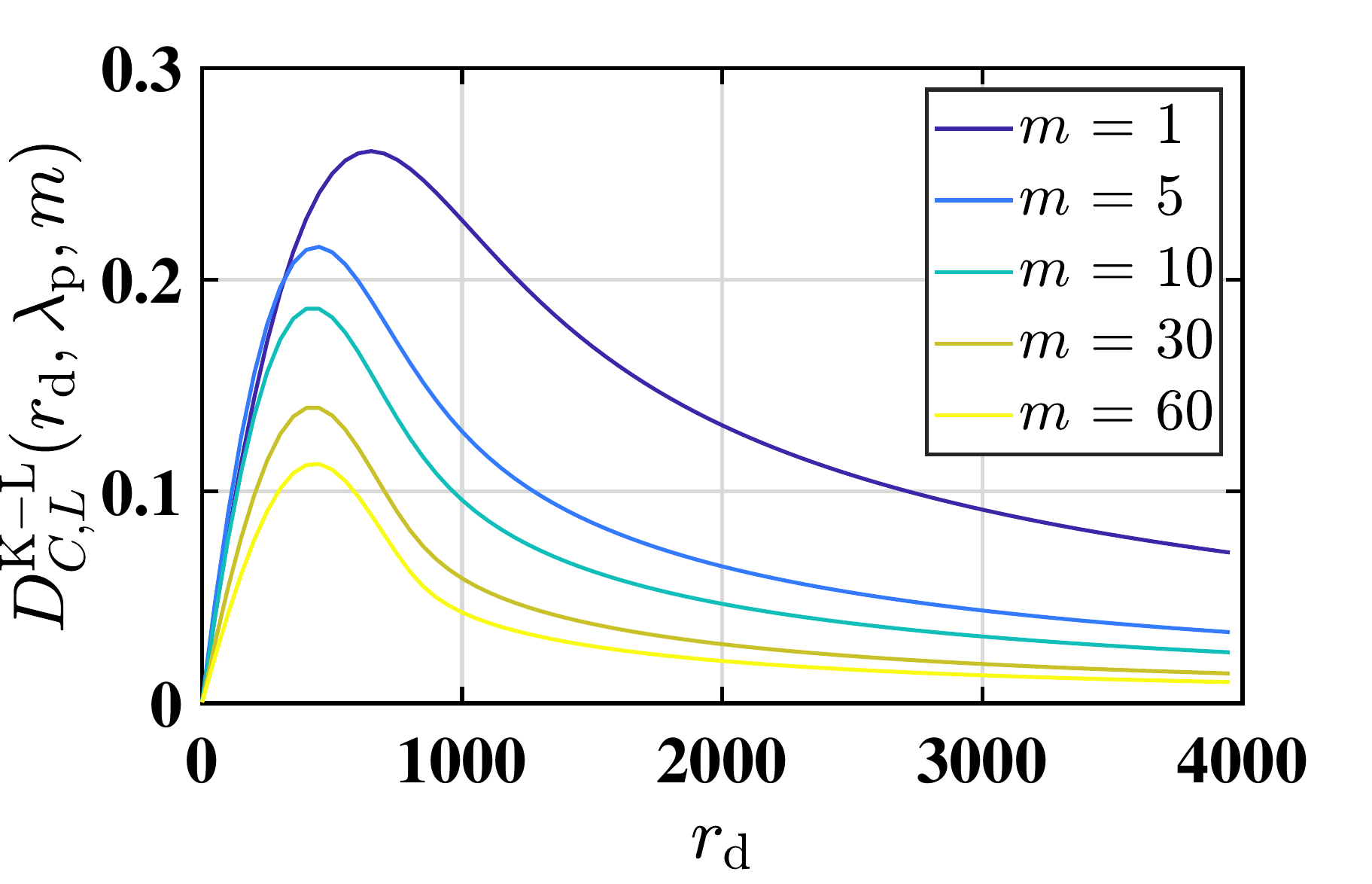}}
	\hfill
	\caption{\small Figures showing the K-S distance, the average  deviation and the K-L distance of bounds from the exact CDF  of contact distance. The intensity of the parent point process $\lambda=10^{-6}$.}\label{UB_analy5_1}
	\hfill
\end{figure}

\textbf{Error bounds for the contact distance:}\\
We now show the variation of the above three metrics for the contact distance for an extensive range of $r_\drm$ and several combinations of $m$. Fig. \ref{UB_analy5_1} shows the maximum deviation (K-S distance), average deviation, and the K-L distance of the bounds from the exact CDF \eqref{eq:FinalMCDN} of the contact distance. 
\begin{figure}[ht!]
	\hfill
	\subfigure[K-S distance of the upper bound]{\includegraphics[width=8cm]{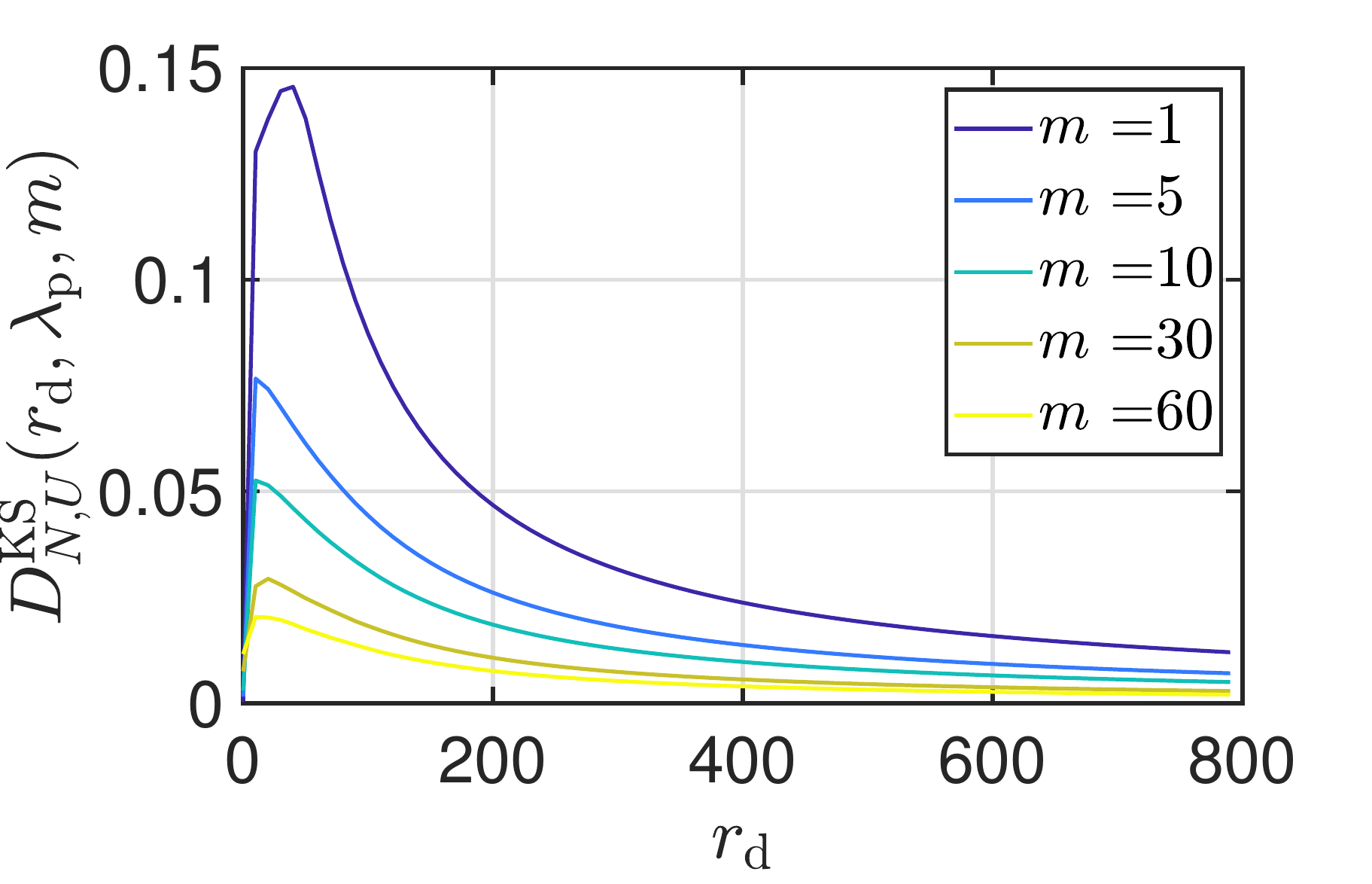}}
	\hfill
	\subfigure[K-S distane of the lower bound]{\includegraphics[width=8cm]{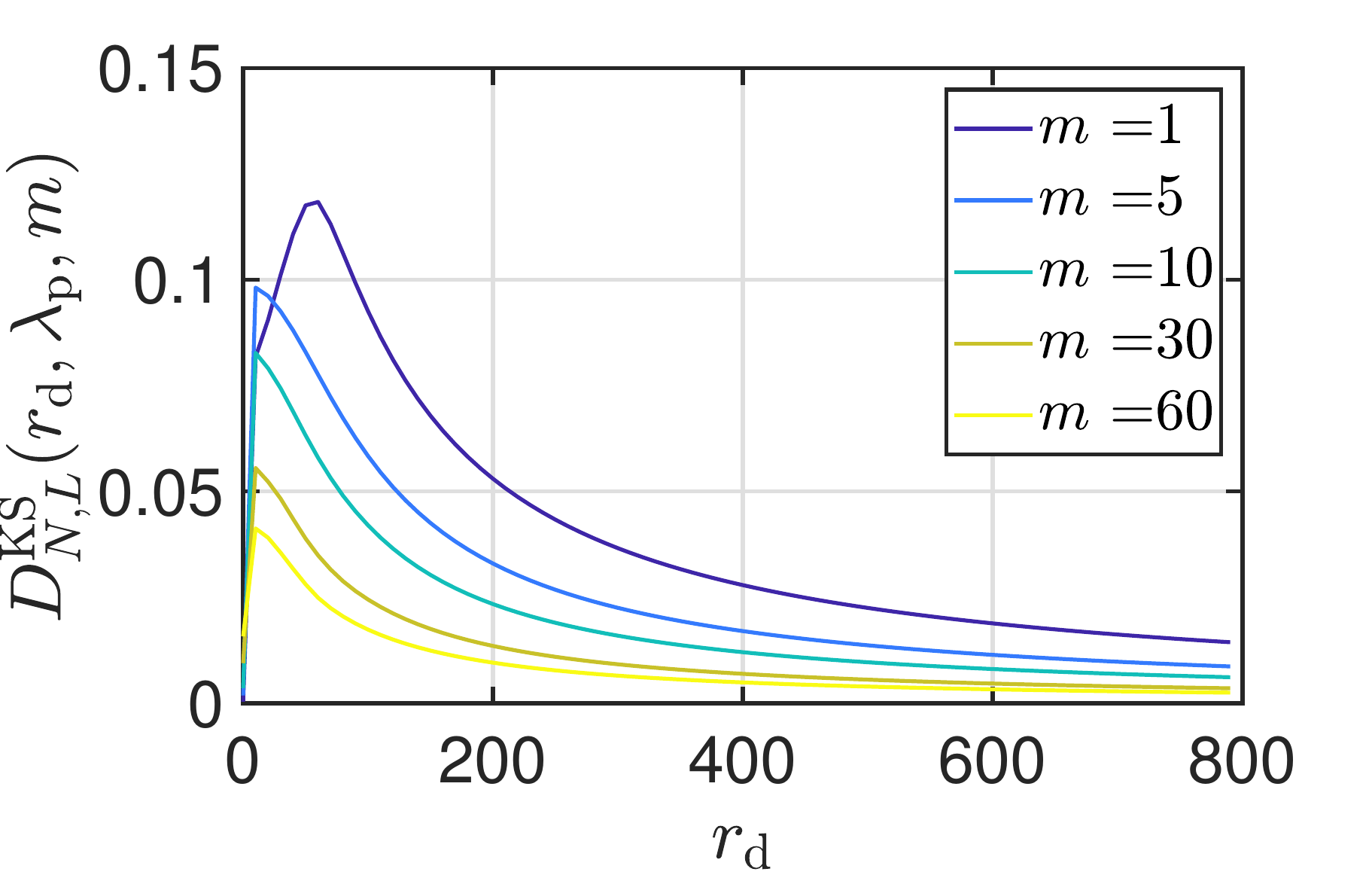}}
	\hfill
	\subfigure[Average deviation of the upper bound]{\includegraphics[width=8cm]{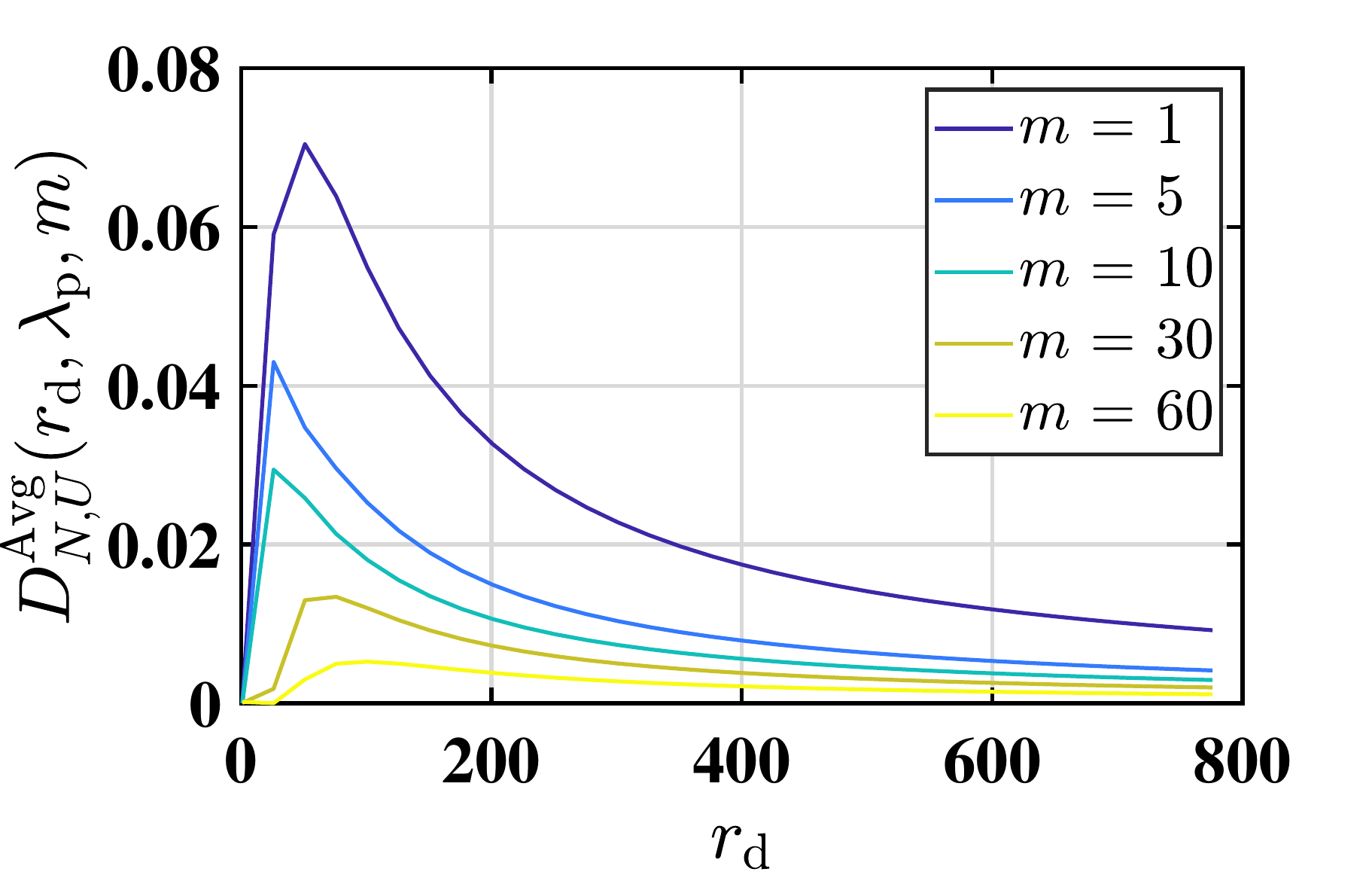}}
	\hfill
	\subfigure[Average deviation of the lower bound]{\includegraphics[width=8cm]{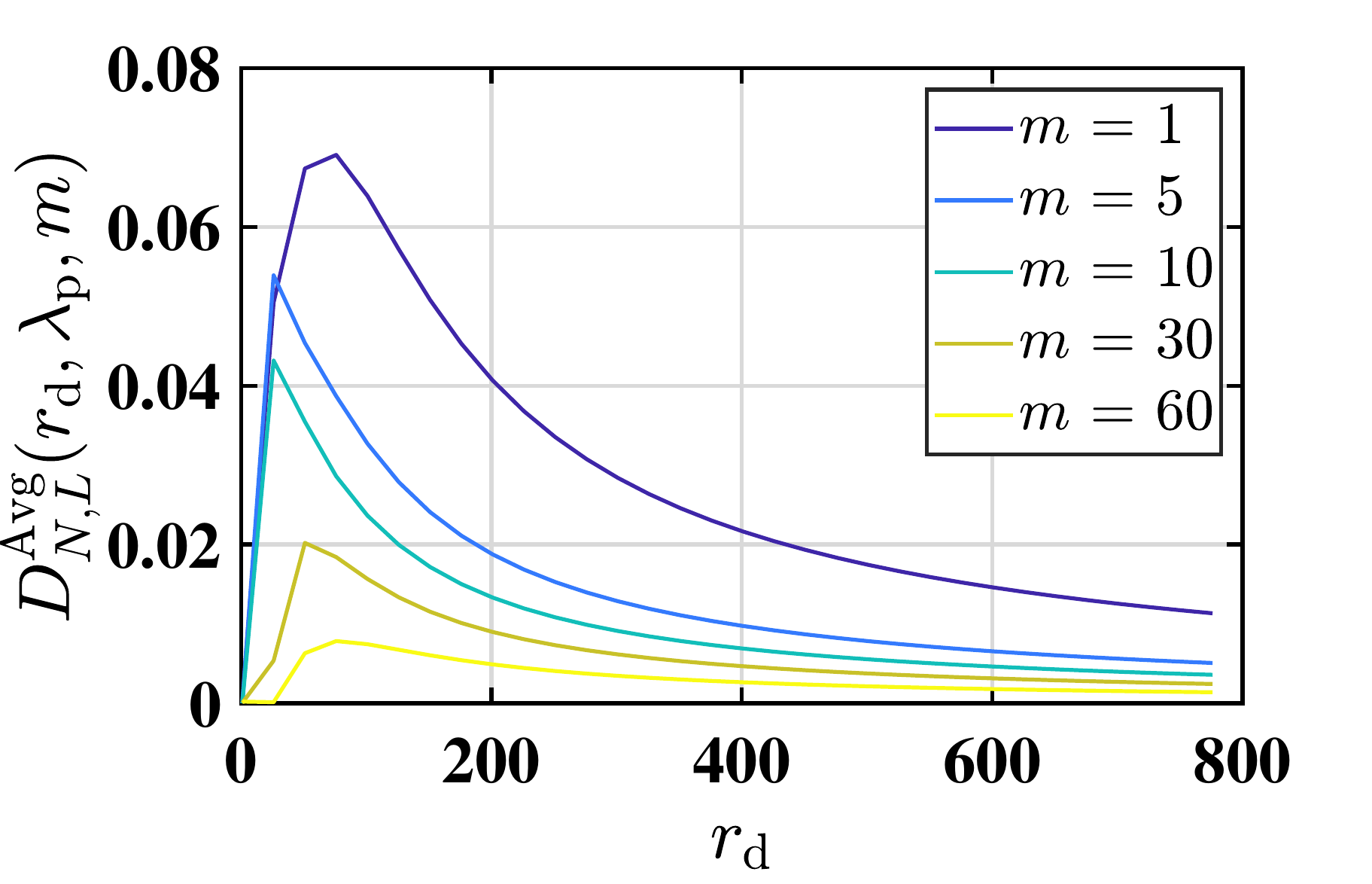}}
	\hfill
	\subfigure[K-L distance of upper bound]{\includegraphics[width=8cm]{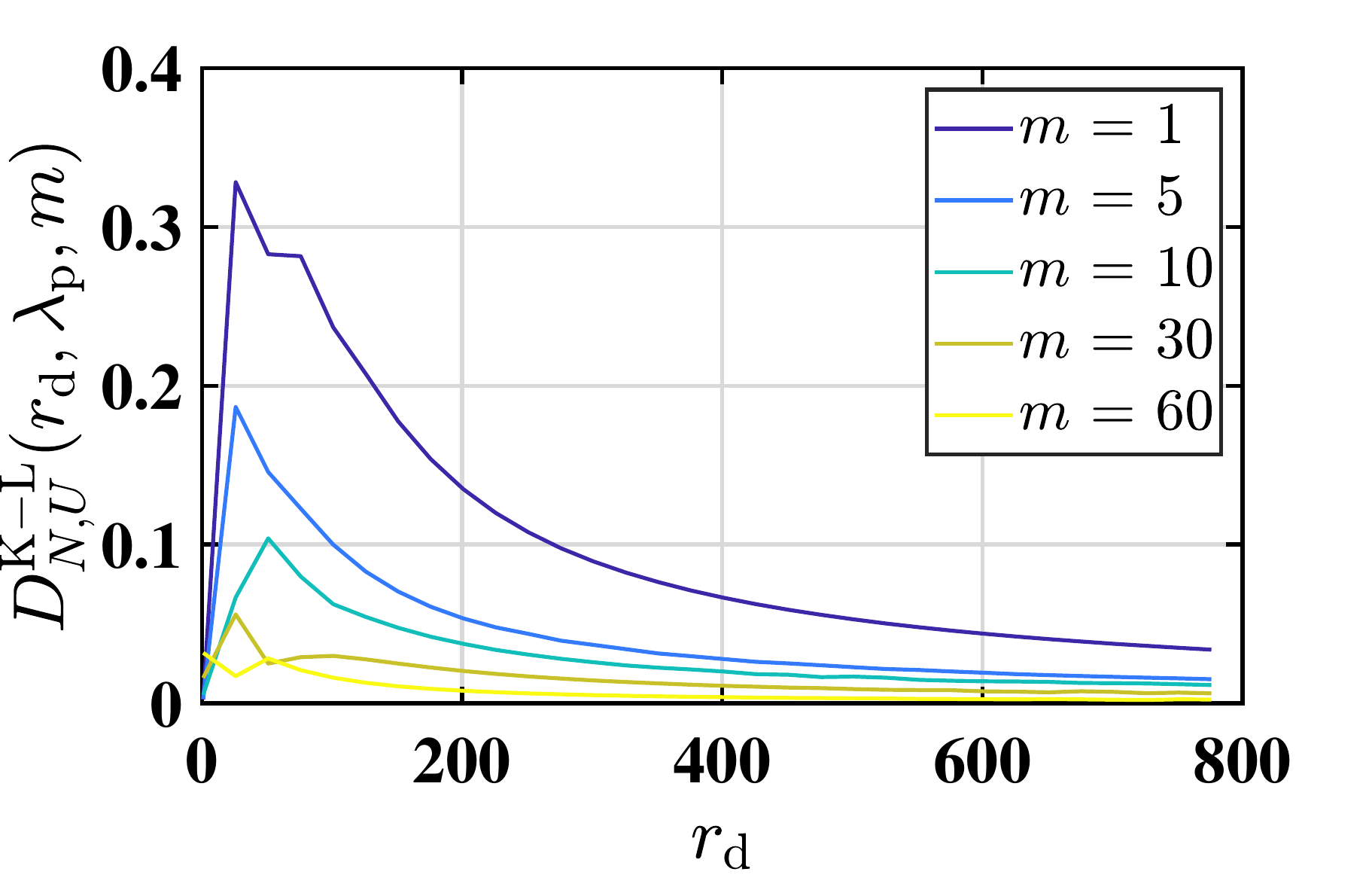}}
	\hfill
	\subfigure[K-L distance of lower bound]{\includegraphics[width=8cm]{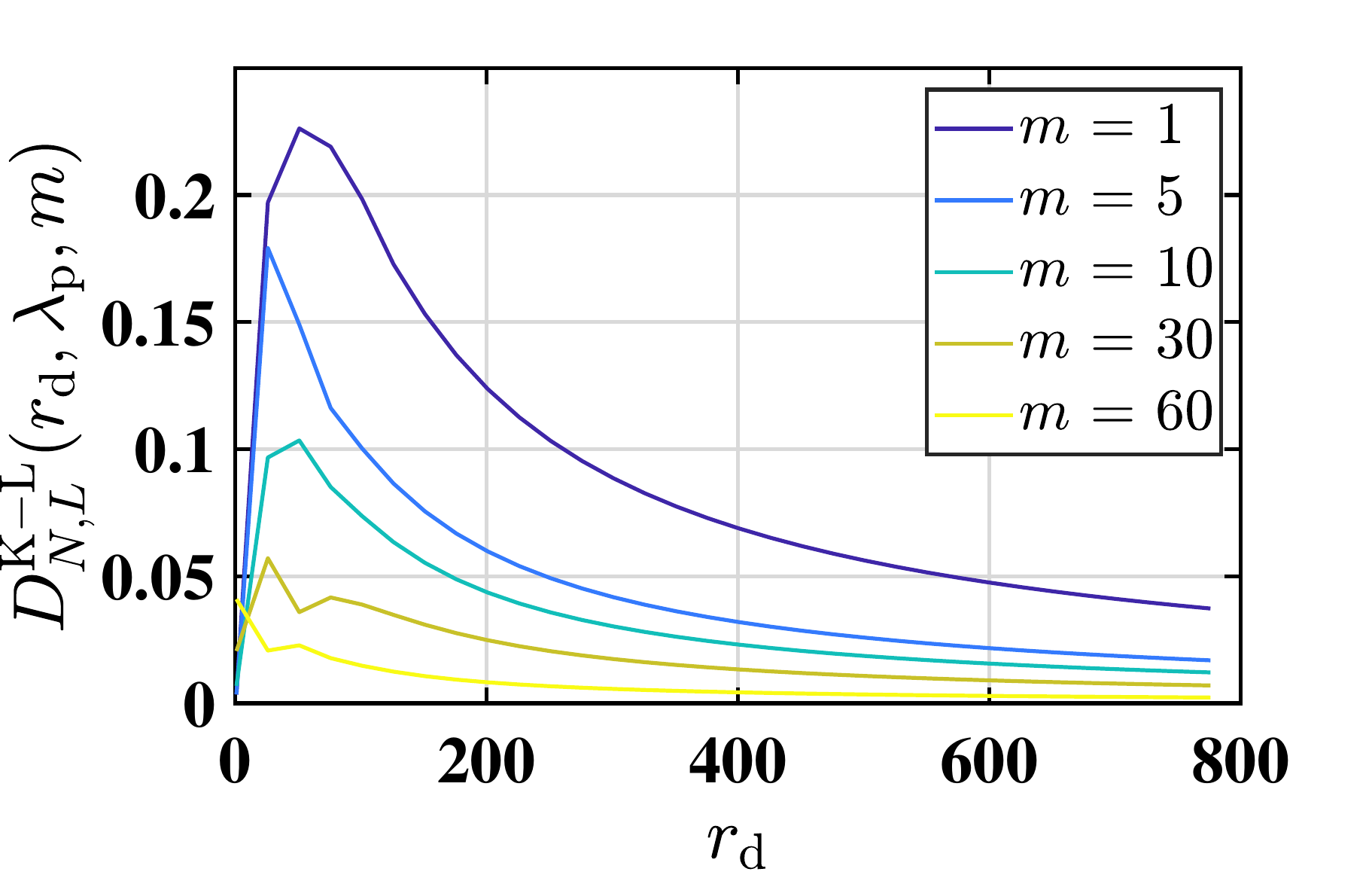}}
	\hfill
	\caption{\small Figures showing the K-S distance, the average deviation  and the K-L distance of bounds from the exact CDF of the nearest neighbor distance $R_{\Nt}$. The intensity of the parent point process $\lambda=10^{-4}$.}\label{UB_analy5_2}
	\hfill
\end{figure}
\begin{enumerate}
	\item \textbf{K-S distance:} For each value of $r_\drm$, we find the K-S distance by considering a large range of $r$. We can observe that the bounds are tight enough for all values of $m$ and the tightness of the bounds improves with the increasing value of $m$.
	We can also observe from the Fig.\ref{UB_analy5_1} (a) and (b) that the  K-S distance curve only shifts  when the intensity of the parent point process $\lambda_{\pt}$ is varied, and hence its behavior remains the same. In particular,  decreasing the intensity  $\lambda_{\pt}$  increases the value of $r_{\drm}$ at which the K-S distance ${D}^{\mathrm{KS}}_{C,\cdot}(r_\drm,\lambda_\pt,m)$ is maximized.
	\item \textbf{Average deviation:} The average value of the deviation of the upper and lower bound is less than $0.09$ for $m=1$ and it decreases further as the value of $m$ is increased. The expected value of deviation reduces significantly with $m$.
	\item \textbf{K-L distance:} Fig. \ref{UB_analy5_1} also shows the K-L distance of the PDF of contact distance with its corresponding upper and lower bound. 
\end{enumerate} 

\begin{figure}[ht!]
	\includegraphics[width=1\columnwidth]{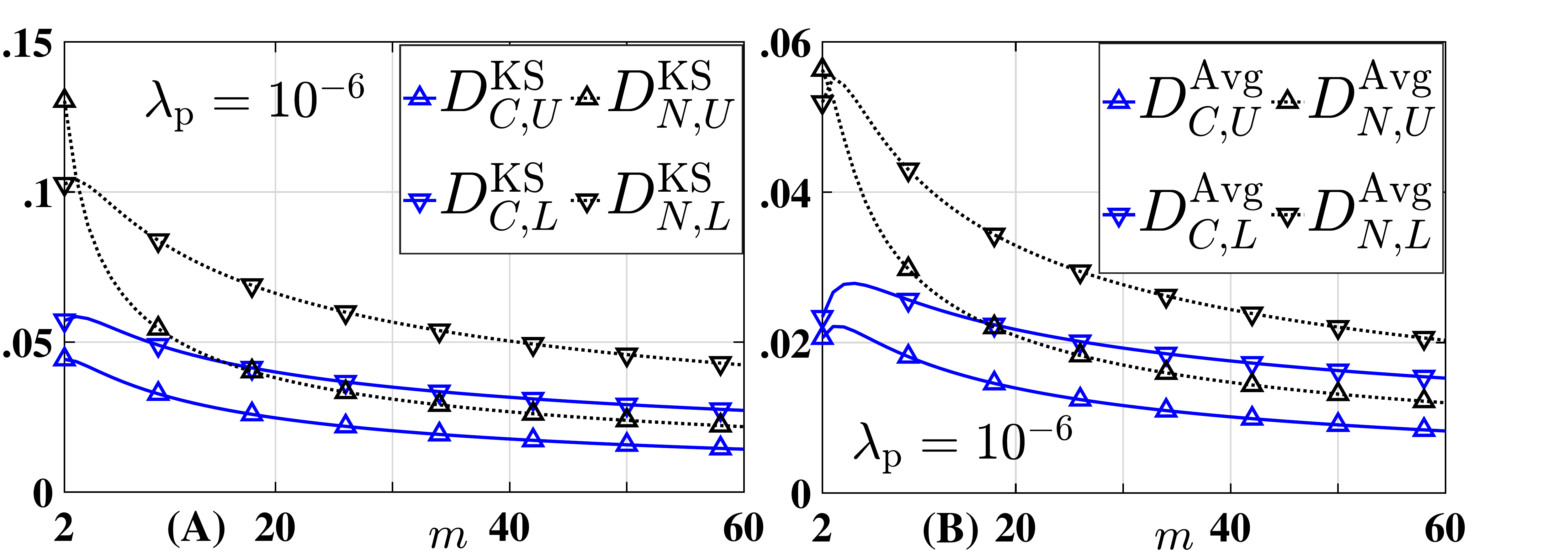}
	\caption{The maximum (over $r_\drm$) of the K-S distance (left figure) and the average deviation (right figure) between proposed bounds and the CDF of CD and NND for different values of $m$. 
		\label{fig:deviation}
	}
\end{figure}

We can observe that the bounds are tight enough for all values of $m$ and the tightness of the bounds improves with the increasing value of $m$. Note that the parameter $m$ represents the extent of clustering. Higher value of $m$ denotes higher order of clustering. We observe that for moderate values of $m$ ({\em i.e.} $m>10$), the proposed bounds have an average deviation ${D}^{\mathrm{Avg}}_{C,\cdot}(r_\drm,\lambda_\pt,m)$ of less than 0.04 for all values of $r_\drm$.  \\
\textbf{Error bounds for the nearest neighbor distance:}
Fig. \ref{UB_analy5_2} shows the K-S distance and the average deviation and the K-L distance of the proposed bounds from the exact expression of CDF. The non-differentiablity in the curves is because of the expression is defined differently in two regions over $r$. We can observe that for  $m>5$, the maximum deviation of the proposed bounds is less than $0.06$  and $0.08$  for  the upper and lower bound respectively. The maximum value of the expected deviation is less than $0.07$, which reflects the tightness of the bounds.  Fig. \ref{UB_analy5_2} also shows the K-L distance of the PDF of the nearest neighbor distance from the bounds.
\begin{figure}[ht!]
	\centering
	\includegraphics[width=\textwidth]{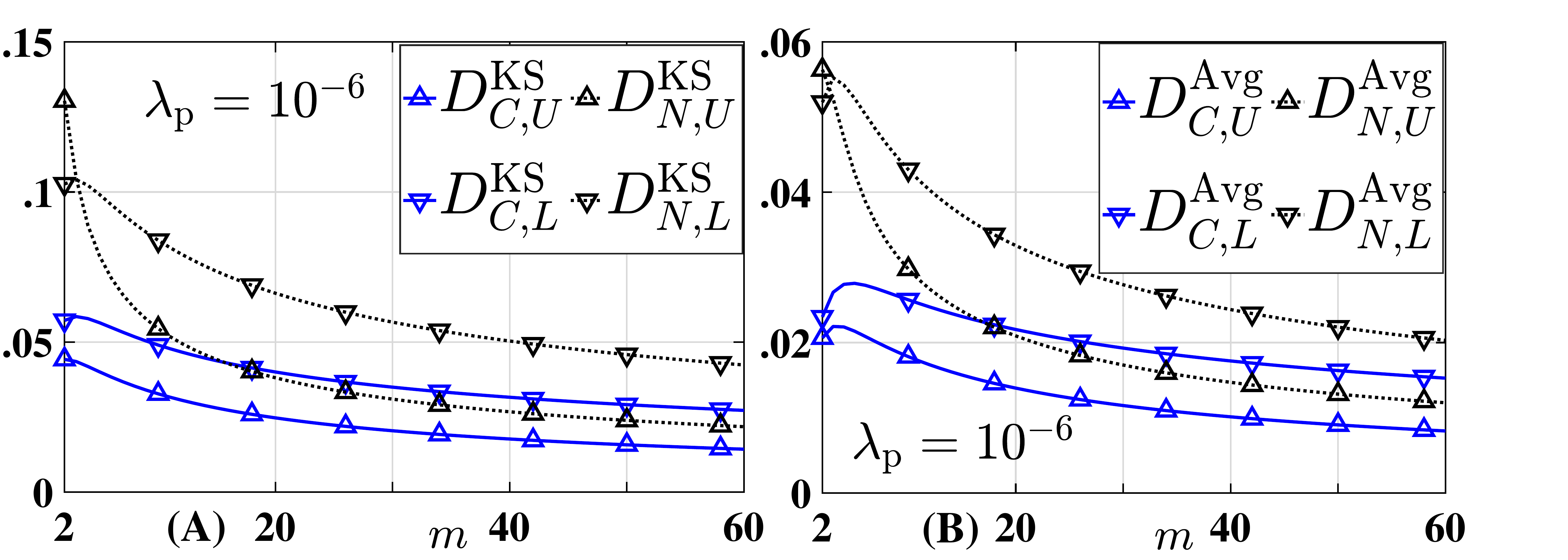}
	\caption{
		The maximum (over $r_\drm$) of the K-S distance (left figure) and the average derivation (right figure) between proposed bounds and the CDF of CD and NND for different values of $m$.
	}\label{CCDF6}
\end{figure}
Fig. \ref{fig:deviation} shows the deviation of bounds. 
${D}^{\cdot}_{C,U}$ and ${D}^{\cdot}_{C,L}$ denote the deviation of $\UBt{F_{R_{\Ct}}}$  and $\LBt{F_{R_{\Ct}}}$ from the exact CDF of CD. ${D}^{\cdot}_{N,U}$ and ${D}^{\cdot}_{N,L}$ denote the deviation of $\UBt{F_{R_{\Nt}}}$  and $\LBt{F_{R_{\Nt}}}$ from  the exact CDF of NND. $\mathrm{K}$$\mathrm{S}$ and $\mathrm{Avg}$ denote the maximum (over $r_\drm$) of the K-S  (Kolmogorov-Smirnov) distance and the average deviation of the respective bound. It can be seen that the proposed bounds are tight. 

\section{Analysis of the PDFs of Contact and Nearest Neighbor Distance}
The PDFs play a crucial role in the analysis of wireless networks. In the supplementary document, we are providing the expressions for the PDF of CD and NND. The PDF $f_{R_\Ct}(r)$ of the contact distance is given as
\begin{align*}
&f_{R_\Ct}(r)=\frac{\dv}{\dv r} F_{R_\Ct}(r).
\end{align*}
Using Leibniz integral rule,
\begin{align*}
f_{R_\Ct}(r)&=n v_n \lambda_{\pt}\exp{\left(-v_n n \lambda_{\pt}
	\int_{0}^{r+r_\drm}\left(1-e^{-\lambda_\drm \A{r}{r_\drm}{x}}\right)x^{n-1}  \dv x\right)}\frac{\dv}{\dv r}\left[\int_{0}^{r+r_\drm}\left(1-e^{-\lambda_\drm \A{r}{r_\drm}{x}}\right)x^{n-1}\right]\\
&=n v_n \lambda_{\pt}\exp{\left(-v_n n \lambda_{\pt}
	\int_{0}^{r+r_\drm}\left(1-e^{-\lambda_\drm \A{r}{r_\drm}{x}}\right)x^{n-1}  \dv x\right)}\left[\left(1-e^{-\lambda_\drm \A{r}{r_\drm}{r+r_\drm}}\right)(r+r_\drm)^{n-1}\right.\\
&\left.+\int_{0}^{r+r_\drm}\frac{\dv}{\dv r}\left(1-e^{-\lambda_\drm \A{r}{r_\drm}{x}}\right)x^{n-1}\dv x\right]\\
&=n v_n \lambda_{\pt}\exp{\left(-v_n n \lambda_{\pt}
	\int_{0}^{r+r_\drm}\left(1-e^{-\lambda_\drm \A{r}{r_\drm}{x}}\right)x^{n-1}  \dv x\right)}\left[\vphantom{\frac{e^2}{5}}\left(1-e^{-\lambda_\drm \A{r}{r_\drm}{r+r_\drm}}\right)(r+r_\drm)^{n-1}\right.\\
&\left.+\lambda_\drm\int_{0}^{r+r_\drm}\left(\frac{\dv}{\dv  r}[ \A{r}{r_\drm}{x}]\right)\left( e^{-\lambda_\drm \A{r}{r_\drm}{x}}\right)x^{n-1}\dv x\right].
\end{align*}
Note that the area of intersection of the balls that are $r+r_\drm$ distance apart is zero {\em i.e.} $\A{r}{r_\drm}{r+r_\drm}=0$. Putting this value in the expression of the PDF of CD, we get the simplified expression as:
\begin{align}
f_{R_\Ct}(r)&=nv_n\lambda_{\pt}\lambda_{\drm}\exp{\left(-v_n n \lambda_{\pt}
	\int_{0}^{r+r_\drm}\left(1-e^{-\lambda_\drm \A{r}{r_\drm}{x}}\right)x^{n-1}  \dv x\right)}\times\nonumber\\&\int_{0}^{r+r_\drm}\left(\frac{\dv}{\dv r}[ \A{r}{r_\drm}{x}]\right)\left( e^{-\lambda_\drm \A{r}{r_\drm}{x}}\right)x^{n-1}\dv x
\end{align}  
\begin{figure}[ht!]
	\centering
	\includegraphics[width=.5\columnwidth]{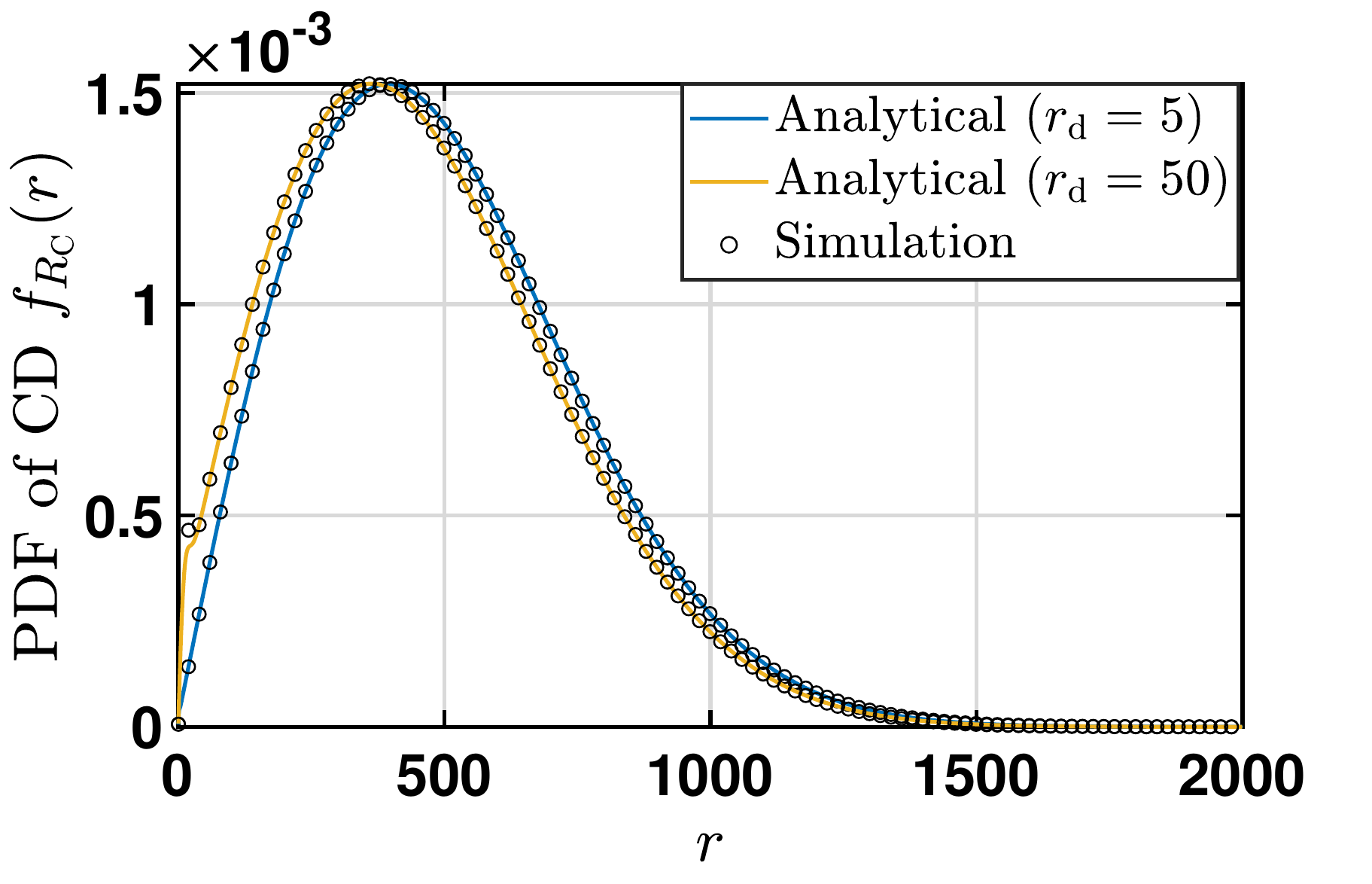}
	\caption{PDF for the contact distance}
	\label{PCDF4}
\end{figure} 
Fig.\ref{PCDF4} shows the PDF of CD.

Similarly, we can also derive the PDF of the nearest neighbor distance. The CDF of NND for $r\leq2r_\drm $ is:
\begin{align*}
F_{R_\Nt}(r)=1-(1-F_{R_\Ct}(r))\frac{n}{r_{\drm}^{n}} \int_{0}^{r_{\drm}}e^{
	-\lambda_\drm
	\A{r}{r_\drm}{x}} 
x^{n-1}\dv x.
\end{align*}
The PDF of nearest neighbor, for the case $r\leq2r_\drm $, is:
\begin{align*}
f_{R_\Nt}(r)&=\frac{nf_{R_\Ct}(r)}{r_\drm^n}\int_{0}^{r_\drm}e^{-\lambda_\drm\A{r}{r_\drm}{x}}x^{n-1}\dv x\\
& +\frac{n\lambda_\drm(1-F_{R_\Ct}(r))}{r_{\drm}^{n}}\left[\int_{0}^{r_\drm}\left(\frac{\dv}{\dv r}\A{r}{r_\drm}{x}\right)e^{-\lambda_\drm\A{r}{r_\drm}{x}}x^{n-1}\dv x\right].
\end{align*}
The expression of the CDF, for $r>2r_\drm$, is
\begin{align*}
F_{R_\Nt}(r)=1-(1-F_{R_\Ct}(r))e^{-m}.
\end{align*}
Hence, the PDF of NND, for $r>2r_\drm$, is given as
\begin{align*}
f_{R_\Nt}(r)=f_{R_\Ct}(r)e^{-m}.
\end{align*}

\bibliographystyle{ieeetran}

\end{document}